\documentclass[11pt]{article}

\usepackage{jheppub}
 
\usepackage{graphicx, epsfig} %include figure files
\usepackage{subfig}
\usepackage{amsmath,amssymb,amsfonts,dsfont,mathrsfs,amsthm,mathtools}
\usepackage{bm} %include bold math: \bm{} creates bold letters in math mode
\usepackage{color}
\usepackage[usenames]{xcolor}
\usepackage{hyperref}
\usepackage{siunitx}
\hypersetup{colorlinks=true,urlcolor=blue,linkcolor=magenta,citecolor=blue,filecolor=blue}
\usepackage[normalem]{ulem}
\usepackage{array}
\usepackage{hyperref}
\usepackage{booktabs}
\usepackage{blindtext}
\usepackage{upgreek}
\usepackage{subfloat}
\usepackage[normalem]{ulem}
\newcommand{\stkout}[1]{\ifmmode\text{\sout{\ensuremath{#1}}}\else\sout{#1}\fi}

\newcommand{\diff}[1]{\text{d}#1}

\newcommand{\GN}{G_\text{N}}
\newcommand{\SEE}{S_{\text{EE}}}
\newcommand{\Ls}{L_\star}
\newcommand{\RT}{{\Sigma_{A}}}
\newcommand{\cl}{{\Sigma_{\text{cl}}}}
\newcommand{\ext}{{\Sigma_{\text{ext}}}}
\newcommand{\IE}{I_{\text E}}
\newcommand{\IEren}{I_{\text E}^{\text{ren}}}

\newcommand{\ICG}{I_{\text{CG}}}
\newcommand{\detg}{\sqrt{|g|}}
\newcommand{\deth}{\sqrt{|h|}}

\makeatletter
\newcommand{\dal}{\mathop{\mathpalette\dal@\relax}}
\newcommand{\dal@}[2]{%
  \begingroup
  \sbox\z@{$\m@th#1\square$}%
  \dimen0=\fontdimen8
    \ifx#1\displaystyle\textfont\else
    \ifx#1\textstyle\textfont\else
    \ifx#1\scriptstyle\scriptfont\else
    \scriptscriptfont\fi\fi\fi3
  \makebox[\wd\z@]{%
    \hbox to \ht\z@{%
      \vrule width \dimen0
      \kern-\dimen0
      \vbox to \ht\z@{
        \hrule height \dimen0 width \ht\z@
        \vss
        \hrule height 2\dimen0
      }%
      \kern-2.5\dimen0
      \vrule width 2.5\dimen0
    }%
  }%
  \endgroup
}
\makeatother

\begin{document}
%%%%%%%
\title{\huge Higher-dimensional Willmore energy as holographic entanglement entropy}

%%%%%%%

\author[a]{Giorgos Anastasiou,}
\author[b]{Ignacio J. Araya,}
\author[c]{Pablo Bueno,}
\author[d,c]{\\Javier Moreno,}
\author[e]{Rodrigo Olea,}
\author[f]{Alejandro Vilar Lopez}

\affiliation[a]{Universidad Adolfo Iba\~nez, Facultad de Artes Liberales, Departamento de Ciencias,\\ Av. Diagonal Las Torres, 2640 Pe\~nalolen, Chile \vspace{0.1cm}}
\affiliation[b]{Universidad Andres Bello, Departamento de F\'isica y Astronom\'ia, \\  Facultad de Ciencias Exactas, Sazi\'e 2212, Piso 7, Santiago, Chile \vspace{0.1cm}}
\affiliation[c]{Departament de F\'isica Qu\'antica i Astrof\'isica, Institut de Ci\'encies del Cosmos,\\
Universitat de Barcelona, Mart\'i i Franqu\'es 1, E-08028 Barcelona, Spain \vspace{0.1cm}}
\affiliation[d]{Departamento de F\'isica, Universidad de Concepci\'on, Casilla, 160-C, Concepci\'on, Chile \vspace{0.1cm}}
\affiliation[e]{Instituto de Física, Pontificia Universidad Católica de Valparaíso, Casilla 4059, Valparaíso, Chile.}
\affiliation[f]{Physique Th\'eorique et Math\'ematique and International Solvay Institutes,\\
Universit\'e Libre de Bruxelles (ULB), C.P. 231, 1050 Brussels, Belgium \vspace{0.1cm}}

\vspace{0.3cm}
\emailAdd{georgios.anastasiou@uai.cl} 
\emailAdd{ignacio.araya@unab.cl}
\emailAdd{pablobueno@ub.edu}
\emailAdd{fjaviermoreno@udec.cl}
\emailAdd{rodrigo\_olea\_a@yahoo.co.uk}
\emailAdd{alejandro.vilar.lopez@ulb.be}

\abstract{
The vacuum entanglement entropy of a general conformal field theory (CFT) in $d=5$ spacetime dimensions contains a universal term, $F(A)$, which has a complicated and non-local dependence on the geometric details of the region $A$ and the theory. Analogously to the previously known $d=3$ case, we prove that for CFTs in $d=5$ which are holographically dual to Einstein gravity, $F(A)$ is equal to a four-dimensional version of the ``Willmore energy'' associated to a doubled and closed version of the Ryu-Takayanagi (RT) surface of $A$ embedded in $\mathbb{R}^5$. This generalized Willmore energy is shown to arise from a conformal-invariant codimension-two functional obtained by evaluating six-dimensional Conformal Gravity on the conically-singular orbifold of the replica trick. The new functional involves an integral over the doubled RT surface of a  linear combination of three quartic terms in extrinsic curvatures and is free from ultraviolet divergences by construction. We verify explicitly the validity of our new formula for various entangling regions and argue that, as opposed to the $d=3$  case, $F(A)$ is not globally minimized by a round ball $A=\mathbb{B}^4$. Rather, $F(A)$ can take arbitrarily positive and negative values as a function of $A$. Hence, we conclude that the round ball is not a global minimizer of $F(A)$ for general five-dimensional CFTs.
}

\maketitle
\section{Introduction}
%\comment{vacuum EE captures universal information (all CFT data should be obtainable from knowledge of EE for general regions); general comments on shape dependence; }\\

In the context of algebraic quantum field theory \cite{Haag:1963dh}, the entanglement entropy (EE) of spacetime regions   provides a canonical measure of statistical properties of the vacuum state restricted to the algebras attached to those regions. Similarly to vacuum expectation values of local operators in the standard approach \cite{Wightman:1956zz}, it is reasonable to expect that a full characterization of a given theory should be achievable from the knowledge of the vacuum EE of arbitrary regions \cite{Agon:2021lus,Agon:2021zvp,Long:2016vkg, Chen:2016mya, Chen:2017hbk,Calabrese:2010he,Agon:2015ftl,Agon:2024zae,Agon:2022efa}. Of course, the EE is not well-defined in the continuum due to the presence of infinite correlations between fluctuations localized arbitrarily close to the entangling surface --- see \emph{e.g.}, \cite{Haag:1992hx,Witten:2018zxz}. Hence, one is forced to either resort to alternative well-defined measures such as the mutual information \cite{Casini:2006ws,Casini:2015woa} or to regulate the theory by introducing some sort of ultraviolet (UV) regulator. The idea is that some of the terms in the EE expansion in powers of the regulator should be independent of the regulator choice, hence capturing ``universal'' information about the corresponding continuum theory. This is indeed the case, and the EE universal terms have been shown to contain a remarkable amount of information, such as: trace-anomaly coefficients, renormalization group charges, stress-tensor and other conserved currents correlators, thermal entropy charges, conformal bounds involving ratios of some of those quantities, unitarity bounds and more --- see \emph{e.g.}, \cite{Calabrese:2004eu,Calabrese:2009qy,Casini:2011kv, Casini:2015woa,Holzhey_1994, Solodukhin:2008dh, Fursaev:2013fta,Safdi:2012sn,Miao:2015iba,Dowker:2010yj,Perlmutter:2013gua,Faulkner:2015csl,Bueno:2015rda,Bueno:2015qya,Swingle:2013hga,Bueno:2018xqc,Lee:2014xwa,Lewkowycz:2014jia,Casini:2021raa,Anastasiou:2022pzm,Baiguera:2022sao,Huerta:2022cqw,Bueno:2022jbl,Bueno:2023gey}. In characterizing such terms for general quantum field theories in various spacetime dimensions $d$, the interplay between the dependence on the entangling region shape and the one on the theory under consideration turns out to play a crucial role. The situation is rather different depending on whether $d$ is even or odd, as we review next. We focus on the latter case, which will be the one of interest in the present paper.

\subsection{EE in odd dimensions, shape dependence and holography}
For a smooth entangling region $A$ in a general state of an odd-dimensional CFT, the EE admits an expansion in powers of any suitable UV regulator $\delta$ of the form
\begin{equation}\label{eq:Scases}
\SEE(A)=c_{d-2}\left(\frac{H}{\delta}\right)^{d-2}+c_{d-4}\left(\frac{H}{\delta}\right)^{d-4}+\ldots+
c_1\frac{H}{\delta}+(-1)^{\frac{(d-1)}{2}}F(A)\,. \quad \text{(odd }d)    
\end{equation}
Here, the constants $c_{d-2},\dots,c_1$ are all cutoff dependent, $H$ is some characteristic length scale of $A$, and $F(A)$ is a universal, highly non-local and state-dependent constant which captures information about the continuum theory. 

In the vacuum state, and in the particular case of a round ball, $F_0\equiv F\left(A=\mathbb{B}^{d-1}\right)$ coincides with the Euclidean free-energy of the CFT, 
$
F_0=-\log Z_{\mathbb{S}^d}
$ \cite{Dowker:2010yj,Casini:2011kv}. For small deformations of the round ball, the leading correction to $F_0$ is quadratic in the deformation, positive-definite and proportional to the stress-tensor two-point function charge $C_T$ of the corresponding CFT \cite{Mezei:2014zla,Faulkner:2015csl}. As a consequence, the round ball is a local minimum of $F(A)$ for general small deformations of $A=\mathbb{B}^{d-1}$ and for general theories in arbitrary dimensions. A much more challenging question is whether or not it is a \emph{global} minimum, namely, whether or not $F(A)\geq F_0$ holds for arbitrary regions and for general theories. Answering this question is difficult because --- as opposed to the case of even-dimensional CFTs --- the universal term $F(A)$ does not reduce to some combination of fixed theory-independent integrals over the entangling surface controlled by a few theory-dependent coefficients. Rather, in general, the dependence of $F(A)$ on the geometry of $A$ and on the theory under consideration is extremely complicated. In fact,  there are very few known actual models for which an expression for $F(A)$ can be written more or less explicitly for general regions. 

A paradigmatic case is the one of $d$-dimensional holographic CFTs. When the gravitational sector of the bulk theory is given by Einstein gravity with Newton constant $\GN$, the Ryu-Tayakanagi (RT) formula \cite{Ryu:2006bv,Ryu:2006ef,Nishioka:2009un}
\begin{equation}
\SEE(A)= \frac{1}{4 \GN} \mathbf A{\left( \Sigma_A \right)} \,,
\label{RT}
\end{equation}
allows one to compute the EE as the area of the RT surface $\mathbf A\left(\Sigma_A\right)$, a minimal surface in the bulk which is homologous to the entangling region defined in the anti de Sitter (AdS) boundary, where the CFT is defined. Using this, one can obtain expansions of the form (\ref{eq:Scases}) for arbitrary boundary regions, where $\delta$ is a geometric regulator along the holographic direction. With some more work, and in the particular case of $d=3$, it is possible to write an explicit geometric formula for the universal term $F(A)$. In the vacuum state, this reads \cite{Babich:1992mc,Alexakis:2010zz,Fonda:2015nma, Anastasiou:2020smm} %\comment{blah}
\begin{equation}\label{willi3}
F(A)=\frac{\Ls^2}{8\GN}\mathbf W_3\left(2\Sigma_A\right)\, , \quad  \text{where} \quad  \mathbf W_3\left(2\Sigma_A\right)\equiv \frac{1}{4}\int_{2\Sigma_A}\diff^2y\sqrt{\tilde \gamma} \, \tilde{K}^2\, ,
\end{equation}
which holds for holographic Einstein gravity. Here, $\Ls$ is the AdS radius, $2\Sigma_A$ is the surface resulting from taking two copies of the RT surface homologous to $A$ and sewing them together through $\partial A$, and $\tilde \gamma$, $\tilde K$ are, respectively, the induced metric and the trace of the extrinsic curvature $\tilde K_{ab}$ of $2\Sigma_A$ embedded in $\mathbb{R}^3$. The $\mathbf W_3$ functional is well-known in the mathematical literature due to its special properties, and is usually called the ``Willmore energy'' \cite{willmore1965note, willmore1993riemannian}. In particular, it follows straightforwardly that $F(A)/F_0 \geq 1$ for general regions and that the round disk provides a global minimum of $F(A)$. This  holographic result hints at a more general one, namely, that disks globally minimize $F(A)$ for general three-dimensional CFTs. As shown in Ref.~\cite{Bueno:2021fxb} using a combination of geometric arguments and the strong subadditivity property of EE, this is indeed the case, namely,
\begin{equation}\label{fa0}
    F(A)/F_0 \geq 1 \quad \forall\, A \quad \forall\, \text{CFT}_3\, , \quad \text{and} \quad F(A)=F_0 \Leftrightarrow A=\mathbb{B}^{2}\, .
\end{equation}
Additionally, this result selects $F_0$ as a natural normalization for performing comparisons of the EE for different theories. Indeed, it has been recently conjectured that $F(A)/F_0$ for general regions and three-dimensional CFTs is bounded above by the result corresponding to a free scalar field, and below by the one corresponding to a free Maxwell field, giving rise to new sets of conformal bounds \cite{Bueno:2023gey}.\footnote{In particular, applied to the case of slightly deformed disks, this implies that $C_T/F_0 \leq 3/(4\pi^2 \log 2-6\zeta[3])\simeq  0.14887$ for general three-dimensional CFTs. This conjectural universal  bound has been shown to hold for a plethora of theories in Ref.~\cite{Bueno:2023gey}.} %\comment{Hofman-Maldacena-like bounds from EE with important consequences}.

In view of these results, an obvious question arises: what happens in $d=5$? Do all or some of these three-dimensional results possess five-dimensional counterparts? In particular, prior to this paper it is not even know whether or not $F(A)$ has a definite sign in general and whether or not it is bounded from above and/or from below. In order to start addressing these questions, in this paper we generalize Eq.~\eqref{willi3} to five-dimensional holographic theories dual to Einstein gravity. In the vacuum state, we find that the corresponding universal term is in this case given by 
\begin{equation}
F(A)=\frac{\Ls^4}{8\GN}\mathbf W_5\left(2\Sigma_A \right)\, ,
\end{equation}
where the ``generalized Willmore energy'' reads
\begin{equation}\label{ww5}
\mathbf W_5\left(2\Sigma_A\right)\equiv \frac{1}{48}\int_{2\Sigma_A}\diff^4y\sqrt{\tilde\gamma}\left[(\partial\tilde K)^2-\tilde K \tilde K^{ab}\tilde K\tilde K_{ab}+\frac{7}{16}\tilde K^4\right]\, .
\end{equation}
In this case, the doubled RT surface $2\Sigma_A$ is embedded in $\mathbb{R}^5$. Just like in the three-dimensional case, this expression is, by construction, free of UV divergences and can be used to evaluate $F(A)$ for $d=5$ holographic Einstein gravity for general regions once the corresponding  RT surface has been determined. From this expression, which is novel in the physics literature,\footnote{On the other hand, $\mathbf W_5$ has previously been derived using different methods in the mathematical literature. In that context, the construction of $\mathbf W_5$ has included various papers with conflicting results which have finally converged to a functional which agrees with the one presented in Eq.~\eqref{ww5}  --- see Refs.~\cite{guven2005conformally,zhang2017grahamwittens,Graham:2017bew,Gover:2016buc,Blitz:2021qbp,Olanipekun:2021htq,martino2023duality} and references therein.} we can derive a number of general results. On the one hand, it follows that Eq. ~\eqref{fa0} does not go through to the five-dimensional case. Namely, at least for holographic theories, $F(A)$ does not have a sign and, as we show explicitly below, it can take positive, negative and vanishing values for different entangling regions. In particular, while the round ball $\mathbb{B}^4$ remains a local minimum --- which holds true for general theories due to Mezei's formula \cite{Mezei:2014zla,Faulkner:2015csl} --- it is possible to find continuous families of regions for which $F(A)$ first grows as one deforms the ball, then it reaches a maximum, then it vanishes again for some other region and then it takes arbitrarily negative values --- see Figure~\ref{fig:5}. As we explain below, a close look at the five-dimensional free-scalar and free-fermion results for $F(A)$ available in the literature reveals that $F(A)$ does not have a sign in those cases either, so it is reasonable to expect this to be a general feature of $d=5$ CFTs.

Our derivation of the above formula for $\mathbf W_5$ departs from the techniques utilized in Refs.~\cite{Babich:1992mc,Alexakis:2010zz,Fonda:2015nma, Anastasiou:2020smm} for the derivation of $\mathbf W_3$ in the holographic context. It relies on the so-called \emph{Conformal Renormalization} method, which we explain in some detail in the following subsection.

\subsection{Conformal Renormalization and holographic EE}
Because of the geometric properties of asymptotically AdS (AAdS) manifolds, the gravitational on-shell action as well as any other local functional of boundary-anchored hypersurfaces --- such as the holograhpic EE --- are divergent. %, inheriting the volume divergence or equivalently the divergence of the line element at the AdS boundary.
As such, in order to define finite functionals, one needs a renormalization prescription. %as is standard in holography. 
In Refs.~\cite{Henningson:1998gx,deHaro:2000vlm,Bianchi:2001kw,Skenderis:2002wp}, the prescription of holographic renormalization (HR) was developed, wherein a series of boundary counterterms which are defined in terms of the induced metric and its Riemannian curvature are added at the AdS boundary, such that the on-shell gravitational action is rendered finite and its variational principle is made into a well-defined Dirichlet boundary problem. %with respect to the holographic source $g_{(0)ij}$, given by the leading-order coefficient of the asymptotic expansion of the metric near the AdS boundary, which is the canonical conjugate to the holographic stress tensor of the CFT as its response function. 
%These requirements of finiteness and well-defined Dirichlet variational problem are necessary in holography, in order to interpret the gravitational action functional as the generating functional of connected correlators of the CFT, through the GKPW relation in the saddle-point approximation~\cite{Gubser:1998bc,Witten:1998qj}.
Later, in Refs.~\cite{Maldacena:2011mk,Anastasiou:2016jix}, it was shown that the HR counterterms for Einstein-AdS gravity can be obtained asymptotically at the AdS boundary %(a.k.a. conformal boundary)
by embedding the theory into Conformal Gravity (CG). This is possible in four bulk dimensions because every solution of Einstein gravity, with or without a cosmological constant, is also a solution of CG. Moreover, for Einstein-AdS spacetimes, the CG on-shell action is equal to the renormalized on-shell action of Einstein-AdS gravity, expressed in Macdowell-Mansouri form.~\cite{Miskovic:2009bm}. Also, in Ref.~\cite{Grumiller:2013mxa} it was shown that the Weyl-squared action of CG is finite when evaluated for any four-dimensional AAdS manifold. 

In the case of six-dimensional AAdS manifolds, Lü, Pang and Pope (LPP) have shown that there is a unique combination of the three point-wise conformal invariants in six dimensions which admits the Schwarzschild-AdS black hole as a solution \cite{Lu:2011ks,Lu:2013hx}. Interestingly, all Einstein spaces are solutions of this same linear combination of conformal invariants \cite{Anastasiou:2020mik,Anastasiou:2023oro}. We shall refer to this six-dimensional version of CG as ``LPP CG''.
%dubbed LPP Conformal Gravity, or simply Conformal Gravity in 6D. 
Furthermore, when the LPP CG action is evaluated in Einstein-AdS spacetimes, it becomes finite as it reduces to the renormalized Einstein-AdS action~\cite{Anastasiou:2020mik}. This procedure for renormalizing Einstein-AdS gravity by embedding it into CG was dubbed \emph{Conformal Renormalization}.

%At the holographic level, the procedure of embedding Einstein-AdS into conformal gravity was understood in \cite{Maldacena:2011mk} for the 4D CG case, by imposing Neumann boundary conditions for the induced metric at the AdS boundary, which is equivalent to turning off the holographic source which is dual to the partially-massless response ($g_{(1)ij}$ in the asymptotic expansion of the metric). Note that, from the perspective of a counting of degrees of freedom in the bulk theory, CG has both the massless graviton of Einstein-AdS and also a partially massless graviton. The latter is then the bulk degree of freedom whose boundary value corresponds to an additional holographic source~\cite{Grumiller:2013mxa}. The same analysis was done in the 6D CG case, where the boundary condition was identified as an extended Neuman boundary condition where now not only $g_{(1)ij}$ is required to vanish but also $g_{(3)ij}$, which corresponds to an independent holographic source.

In the computation of holographic EE, the Conformal Renormalization prescription provides a natural way to isolate the finite term in odd-dimensional CFTs dual to Einstein-AdS. This is because the EE can be computed directly from the gravitational on-shell action, using the generalized gravitational entropy formula~\cite{Lewkowycz:2013nqa}. The finiteness of the latter gets inherited by the former. Following this idea, in Ref.~\cite{Anastasiou:2022ljq} the holographic EE functional for Einstein-AdS gravity in four bulk dimensions was derived starting from CG. This is achieved by applying the generalized gravitational entropy formula to the CG action, which is evaluated on the conically singular orbifold obtained via the replica trick~\cite{Calabrese:2004eu} and using the relations given in Ref.~\cite{Fursaev:2013fta}. Then, the resulting functional was identified with the integrand of the Graham-Witten anomaly~\cite{Graham:1999pm}, which corresponds to a pointwise conformally invariant functional defined on the codimension-two hypersurface localized at the conical singularity. This functional %, $L_\Sigma$,
was explicitly used to derive not only the renormalized holographic EE of Einstein-AdS, but also the so-called ``reduced Hawking mass'' and Willmore energy functionals which, in other contexts, are related to interesting quantities such as the entanglement susceptibility and to global bounds on information~\cite{Fonda:2015nma}.

For the computation of the renormalized holographic EE for CFTs dual to Einstein-AdS in six bulk dimensions, one expects that the finite part could be obtained directly starting from the holographic EE functional of the LPP CG, as it is the latter action which reduces to the renormalized Einstein-AdS action when evaluated on Einstein manifolds. In the mathematics literature on conformal invariants, the functional which corresponds to the area anomaly of an extremal codimension-two boundary anchored hypersurface in seven-dimensional asymptotically hyperbolic Einstein manifolds is the Graham-Reichert energy~\cite{Graham:2017bew}. As we will see in this work, this functional, up to boundary terms that will be completely fixed for Einstein spacetimes, defines a codimension-two four-dimensional conformal invariant which gives rise to the finite (renormalized) part of the holographic EE for Einstein-AdS gravity in six bulk dimensions and, as a consequence, also defines a well-motivated version for a higher-dimensional generalization of the Willmore energy --- namely, $\mathbf W_5$ as defined in Eq.~\eqref{ww5}. We also provide numerous examples, by explicit computation, that this generalized Willmore energy matches the finite part of the RT functional, and also that it can be directly computed by considering the covariant version of the renormalized holographic EE that is directly obtained from the Graham-Reichert energy (with fixed boundary terms) when evaluated for Einstein manifolds. 

The remainder of the paper is as follows. In Sec.~\ref{sec:2} we review the derivation of Willmore energy $\mathbf W_3$ and reduced Hawking mass $\mathbf I_3$ from evaluating the four-dimensional CG action in a manifold with a conical defect. This recipe generates a codimension-two functional $\mathbf L$, known as Graham-Witten anomaly, from which $\mathbf W_3$ and $\mathbf I_3$ appear as particular cases. In Sec.~\ref{sec:3} we extend the derivation presented in the previous section to the six-dimensional LPP CG action, which, after evaluating in the conically singular manifold, produces a four-dimensional functional $\mathbf F$, which coincides with the Graham-Reichert anomaly. We observe that the analogous particular cases to the ones considered in two dimensions less allow us to identify a generalized Willmore energy $\mathbf W_5$ which matches proposals in the mathematical literature as well as a ``generalized reduced Hawking mass'' $\mathbf I_5$. We show that the relation between holographic EE in $d=3$ and Willmore energy $\mathbf W_3$ by means of the doubling of the RT surface also holds for holographic EE in $d=5$ and the novel $\mathbf W_5$ functional. We perform explicit checks of this relation for entangling regions consisting of round balls, slightly deformed balls and strips, obtaining the same results from both expressions. In Sec.~\ref{sec4} we study the global shape-dependence of the $\mathbf W_5$ functional %employing ellipsoids 
and observe that this quantity is neither bounded from below nor from above. Related observations in the case of free fields lead us to conjecture that $F(A)$ is unbounded both from below and from above for general five-dimensional CFTs. In Sec.~\ref{sec:4} we conclude with some general comments and future directions in light of the results obtained in this paper. Our notation and conventions are summarized in Appendix \ref{conventions}. Certain intermediate calculations explained in the bulk of the paper appear in Appendices \ref{confcomple}, \ref{Appendixcomputation} and \ref{appdiv}.

\section{Holographic EE in $d=3$ as Willmore energy}\label{sec:2}
As a warm up, in this section we review the previously known fact that the holographic EE in the vacuum of three-dimensional CFTs dual to Einstein gravity contains a universal term $F(A)$ which can be written as a Willmore energy associated to a doubled version of the corresponding RT surface. The derivation presented here relies on the holographic renormalization of energy functionals in the context of four-dimensional CG, which we will later extend in the following section to the six-dimensional case.

%The close relation between quantum information theoretical measures and geometry, provided by holography in the form of the Ryu-Takayanagi (RT) relation, opened up a new window in the exploration of properties of EE starting from geometric features of minimal surfaces. An interesting feature along this line of thought, is the fact that the finite part of holographic entanglement entropy of a given spatial region $F(A)$ in a $2+1$ dimensions vacuum state can be computed in terms of the so-called Willmore energy. Let us elaborate in this statement. 

Generally speaking, given a closed smooth two-dimensional surface with genus $g$ embedded in $\mathbb{R}^3$, $\Sigma_g\hookrightarrow\mathbb{R}^3$, its Willmore energy functional is defined as \cite{willmore1965note, willmore1993riemannian}
\begin{equation}\label{eq:Wdef}
\mathbf W_3\left(\Sigma_g\right)\equiv \int_{\Sigma_g} H^2\diff S\, ,
\end{equation}
where $H$ is the mean curvature\footnote{For surfaces embedded in three dimensions, the mean curvature is related to the extrinsic curvature as $2H=K$.} of $\Sigma_g$ and $\diff S$ is the surface element. This quantity has been the subject of intensive study in the mathematical literature because of the existence of general bounds satisfied for arbitrary surfaces $\Sigma_g$. Notably, the Willmore energy for any closed surface embedded in $\mathbb{R}^3$ satisfies
\begin{equation}\label{eq:boundW}
\mathbf W_3\left(\Sigma_g\right)\geq 4\pi\, .
\end{equation}
The inequality is saturated in the case of the round sphere $\Sigma_g=\mathbb{S}^2$ --- \emph{i.e.}, the round sphere is the Willmore energy minimizer among all possible closed surfaces. Restricted to the case of toroidal closed surfaces, $g=1$, the bound is saturated by the so-called ``Clifford torus'', and it can be improved to $\mathbf W_3\left(\Sigma_{g}\right)\geq 2\pi^2$, which holds for $g\geq 1$ \cite{marques2014min}.

The link between holographic EE and Willmore energy can be understood from the prescription described in the following subsections. First, consider the RT surface $\RT$ associated to the entangling region $A$, this is, possessing the same boundary $\partial A=\partial \RT$, homologous and with minimal area. Now, glue an identical copy $\RT'$ along its boundary, \emph{i.e.}, $\partial \RT=\partial \RT'$. The obtained doubled-copied submanifold $2\RT=\RT\cup\RT'$ is a closed surface embeddable (through a Weyl rescalling of the ambient space) in $\mathbb{R}^3$ to which we can associate a Willmore energy\footnote{This relation only works for entangling regions in the ground state of the CFT, which is dual to pure AdS. As the Poincaré patch of a constant-time slice of AdS$_4$ is conformally equivalent to $\mathbb{R}^3/\mathbb{Z}_2$, the procedure considers doubling the bulk across the boundary, and then performing a conformal transformation to obtain $\mathbb{R}^3$. The RT surface is therefore also doubled.} --- in Figure~\ref{fig:2} below the doubling of the RT surface is presented in the case of a spherical entangling surface. Based on this, the finite piece of EE can be expressed as \cite{Babich:1992mc,Alexakis:2010zz,Fonda:2015nma, Anastasiou:2020smm}
\begin{equation}\label{eq:FAW}
    F(A)=\frac{\Ls^2}{8\GN}\mathbf W_3\left(2\RT\right)\, .
\end{equation}
Due to the existence of the aforementioned bounds, the Willmore energy is particularly useful when studying global properties regarding the shape dependence of holographic EE. For instance, the bound \eqref{eq:boundW} allows to establish that the disk entangling region, which corresponds to a spherical surface in the double-copied RT surface $2\RT=\mathbb{S}^2$, minimizes $F(A)$ among all possible smooth shapes. In other words,
\begin{equation}\label{Fminim}
F(A)\geq\frac{\pi L_{\star}^2}{2\GN}\, , \quad  \text{with} \quad F(A)=\frac{\pi L_{\star}^2}{2\GN} \Leftrightarrow A = \text{disk}\, .
\end{equation}
In particular, the shape dependence of $F(A)$ is encoded in the AdS curvature of the RT surface, which admits an upper bound due to the previous relation \cite{Anastasiou:2020smm}. The fact that the round disk minimizes $F(A)$ across all possible entangling regions in the ground state was later proven true for general three-dimensional CFTs in Ref.~\cite{Bueno:2021fxb}. For other applications of Willmore energy in the context of EE in three and four-dimensional CFTs (holographic or not) see Refs.~\cite{Astaneh:2014uba,Perlmutter:2015vma,Seminara:2018pmr,Taylor:2020uwf}.

\subsection{Energy functionals from CG in four dimensions}\label{sec:21}
As we have anticipated in the previous subsection, the Willmore energy, $\mathbf W_3$, captures the universal contribution to the vacuum holographic EE for Einstein gravity in three (boundary) dimensions. As argued in Ref.~\cite{Anastasiou:2022ljq}, $\mathbf W_3$  belongs to a broader class of energy functionals which exhibit restricted conformal symmetry under Weyl rescalings of the ambient metric, alongside the \emph{renormalized area}, $\mathbf{A}^{\text{ren}}$, and the \emph{reduced Hawking mass}, $\mathbf{I}_3$.
%, along with the renormalized area $\mathbf{A}^{\text{ren}}$ and the reduced Hawking mass $\mathbf{I}_3$ that exhibit restricted conformal symmetry under Weyl rescalings of the ambient metric.
In this work we are mostly interested in the first two objects, which are related to the finite piece of the holographic EE.  %$F(A)$ for three-dimensional CFTs dual to Einstein gravity. 
The third one is an interesting byproduct of our analysis that provides information regarding bounds that EE has to satisfy for generic states of a $\left(2+1\right)$ dimensional CFT~\cite{Fischetti:2016fbh}. All these functionals will emanate  from another, which we denote $\mathbf{L}$, defined for codimension-two surfaces embedded in four-dimensional space.

The key input here is the Lewkowycz-Maldacena (LM) prescription  \cite{Lewkowycz:2013nqa} that identifies the generalized gravitational entropy with the holographic EE of the dual CFT. Indeed, the derivation of holographic EE amounts to the evaluation of the Euclidean on-shell action on a singular manifold with conical deficit $2\pi \left (1 -\vartheta \right )$ and differentiating with respect to the angular parameter $\vartheta$. As the angle $\vartheta$ is related to the replica parameter by $\vartheta=1/m$, then, entanglement entropy is obtained in the limit
\begin{equation}\label{eq:LM}
\SEE(A)=-\lim_{\vartheta\rightarrow 1}\partial_\vartheta \IE\left[\mathcal M^{(\vartheta)}_{d+1}\right]\,.
\end{equation}
\sloppy
Namely, there is a one-to-one correspondence between the gravitational action and the codimension-two integral that has to be extremized in order to determine the holographic EE. Based on this consideration one identifies the RT formula as the holographic EE of CFTs dual to Einstein gravity. Indeed, the Ricci scalar contains a conical contribution when evaluated on the orbifold as \cite{Fursaev:1995ef}
% \begin{equation}
%     \frac{1}{16 \pi G_N}\int_{M^{(\vartheta)}}d^4x \sqrt{|g|} R^{(\vartheta)} = \frac{1}{16 \pi G_N} \int_{M}d^4x \sqrt{|g|} R + \frac{(1 - \vartheta)}{4 G_N} \mathcal{A}{\left( \Sigma \right)} \,,
% \end{equation}
\begin{equation}
R\left(\mathcal M^{(\vartheta)} \right)= R\left(\mathcal M\right)+4\pi(1 - \vartheta)\delta_\Sigma\,,
\label{Rconical}
\end{equation}
where $\delta_\Sigma$ is a $(d-1)$-dimensional Dirac delta localized at the conical singularity. In the limit $\theta\rightarrow1$, the RT surface, $\Sigma_A$, is recovered \cite{Dong:2016fnf}, and using Eq.~\eqref{eq:LM} one ends up with the RT formula \eqref{RT}.

In this context, the UV divergences of EE~\eqref{eq:Scases}, or equivalently the area divergences of the RT~\eqref{RT} formula, are identified as the volume divergences of a given gravity action when evaluated on AdS spacetimes. However, the LM prescription~\eqref{eq:LM} suggests that holographic EE functionals coming from renormalized gravitational action, instead of their bare form, are free of UV divergences. Indeed, the authors of Ref.~\cite{Taylor:2016aoi}, inspired by holographic renormalization \cite{Henningson:1998gx,Emparan:1999pm,Balasubramanian:1999re,Kraus:1999di,deHaro:2000vlm,Papadimitriou:2004ap,Papadimitriou:2005ii}, proved that by evaluating the counterterms in the LM formula, one ends up with a series of surface terms that reside at $\partial \RT$ and correctly isolate the universal terms of the holographic EE . However, this prescription does not make manifest certain features of the finite part, such as the conformal invariance of $F(A)$ for vacuum states of three-dimensional CFTs. %, as seen in the last subsection.

An alternative but equivalent path to study these properties is given by \emph{Conformal Renormalization}. This scheme is based on the idea that CG --- a four-derivative gravity theory which is invariant under Weyl rescalings of the metric --- is free of IR divergences for AAdS spacetimes~\cite{Grumiller:2013mxa}, rendering finite any gravitational theory that can be consistently embedded in it, such as Einstein gravity~\cite{Maldacena:2011mk,Anastasiou:2016jix}. We shortly review this connection below.

As it has been shown in Refs.~\cite{Miskovic:2009bm, Anastasiou:2020zwc}, counterterms in four dimensions can be resumed in a unique boundary term with explicit dependence on both the intrinsic and extrinsic curvature \cite{Olea:2005gb,Olea:2006vd,Miskovic:2009bm,Anastasiou:2020zwc}. In particular, this is the case of the second Chern form $\mathcal B_3$, which when added to the four-dimensional Einstein-AdS action 
with the appropriate relative coefficient renders the on-shell action 
\begin{equation}\label{eq:IEren}
\IEren=\frac{1}{16\pi\GN}\int_{\mathcal M}\diff^4 x\detg\, \left(R+\frac{6}{\Ls^2}\right)+\frac{\Ls^2}{64\pi \GN}\int_{\partial \mathcal M}\diff^3 X \mathcal B_3\, ,
\end{equation}
finite. The explicit expression of the second Chern form is given by 
\begin{equation}
\mathcal B_3=-4\deth\, \delta^{\mu_1\mu_2\mu_3}_{\nu_1\nu_2\nu_3}k^{\nu_1}_{\mu_1}\left(\frac{1}{2}r^{\nu_2\nu_3}_{\mu_2\mu_3}-\frac{1}{3}k^{\nu_2}_{\mu_2}k^{\nu_3}_{\mu_3}\right)\, ,    
\end{equation}
%\begin{equation}
%\mathcal B_3=4\deth\, \delta^{\mu_1\mu_2\mu_3}_{\nu_1\nu_2\nu_3}k^{\nu_1}_{\mu_1}\left(\frac{1}{2}r^{\nu_2\nu_3}_{\mu_2\mu_3}-\frac{1}{3}k^{\nu_2}_{\mu_2}k^{\nu_3}_{\mu_3}\right)\, ,
%\end{equation}
where $r_{\mu\nu\rho\sigma}$ is the intrinsic Riemann tensor of $\partial \mathcal M$ and $k_{\mu\nu}$ its extrinsic curvature. In the case of a compact manifold, the boundary term can be traded with quantities defined in the bulk using the Gauss-Bonnet theorem
\begin{equation}\label{eq:GBt}
    \int_{\mathcal M}\diff^4 x\detg\, \mathcal X_4=32\pi^2\chi\left(\mathcal M\right)+\int_{\partial \mathcal M}\diff^3 x\, \mathcal B_3\, ,
\end{equation}
where $\mathcal X_4=R_{\alpha\beta\gamma\delta}R^{\alpha\beta\gamma\delta}-4R_{\alpha\beta}R^{\alpha\beta}+R^2$ is the Gauss-Bonnet term --- or four-dimensional Euler density --- and $\chi(\mathcal M)$ is the Euler characteristic of the manifold $\mathcal M$.

In Ref.~\cite{Miskovic:2009bm} it was shown that the expression for the renormalized Einstein-AdS action \eqref{eq:IEren}, after employing the Gauss-Bonnet theorem, can be recast in the form of the MacDowell-Mansouri action \cite{MacDowell:1977jt} 
\begin{equation}
\IEren=\frac{\Ls^2}{256\pi\GN}\int_{\mathcal M}\diff^4x\detg\, \mathcal Y_4\big|_{\text E}-\frac{\pi \Ls^2}{2\GN}\chi\left(
\mathcal M\right) \,,
\label{reneinsteinads}%
\end{equation}
where\footnote{In the Conformal Renormalizaion literature, the MacDowell-Mansouri term is often written as the monomial $P_4\left(W\big|_{\text{E}}\right)=\mathcal Y_4\big|_{\text{E}} $. This contextualizes the notation employed afterwards in Eq.~\eqref{EH2} when discussing the six-dimensional case.}
\begin{equation}
\mathcal Y_4\big|_{\text E}\equiv \delta_{\alpha_{1} \ldots \alpha_{4}}^{\beta_{1}\ldots \beta_{4}} W_{\beta_{1} \beta_{2}}^{\alpha_{1} \alpha_{2}}\Big|_{\text E}W_{\beta_{3} \beta_{4}}^{\alpha_{3} \alpha_{4}}\Big|_{\text E}\, , \quad \text{and} \quad W_{\alpha\beta}^{\gamma\delta}\Big|_{\text E}=R_{\alpha\beta}^{\gamma\delta}+\frac
{1}{\Ls^2}\delta_{\alpha\beta}^{\gamma\delta} \,,
\label{FAdS}
\end{equation}
is the Weyl tensor for Einstein-AdS spacetimes. The connection to MacDowell-Mansouri comes from the fact that $W\big|_{\text E}$ can be identified as the curvature of the AdS group without torsion.

This notion of curvature suggests a link to a gravity theory where the (full) Weyl tensor plays an essential role, that is, CG. At the fundamental level, the mechanism of embedding an Einstein action supplemented with a topological term in a CG, denoted \textit{conformal covariantization} (c.c.), is far from rigorously defined. However, we will see that it allows to derive the correct codimension-two energy functional. In this line, in order to c.c. $\mathcal Y_4\big|_{\text E}$ we just complete it to full CG, this is
\begin{equation}
    \mathcal Y_4\big|_{\text E}\xrightarrow{\text{c.c.}}\mathcal Y_4\, ,
\end{equation}
so that the resulting CG action reads \cite{Anastasiou:2016jix,Maldacena:2011mk,Anastasiou:2020mik}
\begin{equation}\label{eq:ICG4}
\ICG =\frac{\Ls^2}{64\pi \GN}\int _{\mathcal M}\diff^{4}x\detg\,W^2 -\frac{\pi \Ls ^{2}}{2 \GN}\chi \left (\mathcal M\right )\, ,
\end{equation}
where we denoted $W^2\equiv W_{\alpha\beta\gamma\delta}W^{\alpha\beta\gamma\delta}$. This is based on the fact that, in four dimensions, CG always contains an Einstein sector in its solution set. This sector can be reached upon imposing proper Neumann boundary conditions that eliminate the ghost mode of CG. This result suggests a relation between conformal symmetry in the bulk, realized in the form of CG, and the renormalization of the Einstein-AdS sector, as standard holographic counterterms may be duly reproduced from Eq.~\eqref{reneinsteinads}.

As discussed in Ref.~\cite{Anastasiou:2022ljq}, we can apply Conformal Renormalization to  codimension-two functionals defined in the gravity bulk, making contact with holographic EE. To do so, we evaluate Eq.~\eqref{eq:ICG4} in the orbifold $\mathcal M^{(\vartheta)}$ so that we can employ the LM formula~\eqref{eq:LM} to obtain the EE associated to the region $A$ in the dual CFT. When doing so, additional terms arise from the conical defect for the Euler characteristic $\chi \left (\mathcal M^{(\vartheta)}\right )=\chi \left (\mathcal M\right ) +\left (1 -\vartheta \right )\chi \left (\Sigma \right )$ and the Weyl-squared term \cite{Solodukhin:2008dh, Fursaev:2013fta}
\begin{equation}\label{eq:W2c}
W^2\left (\mathcal M^{(\vartheta)}\right )=W^{2}\left(\mathcal M\right) +8\pi \left (1 -\vartheta \right )\mathcal K_{\Sigma }\,,
\end{equation}
where $\mathcal K_{\Sigma } =R_{AB}^{AB} -R_{A}^{A} +\frac{1}{3}R +\frac{1}{2}K^{2} -{K^A}_{ab} K_A{}^{ab}$ and $K^2=K_AK^A$. Here, $\mathcal K_\Sigma$ is a conformal invariant defined on the codimension-two surface $\Sigma$ to which we associate the metric tensor $\gamma_{ab}$, and the indices $A$, $B$ correspond to the directions  normal to $\Sigma$. Taking this into account, we see that the CG action decomposes, at linear order in $\left(1-\vartheta\right)$, as
\begin{equation}
    \ICG\left (\mathcal M^{(\vartheta)}\right )=\ICG+\frac{\left(1-\vartheta\right)}{4\GN}\, \mathbf{L}(\Sigma)\, ,
\end{equation}
where all the contributions coming from the conical defect are encapsulated in the conformally-invariant codimension-two functional $\mathbf{L}$,\footnote{The $\mathbf{L}$ functional equals the Graham-Witten anomaly \cite{Graham:1999pm}.}
\begin{equation}\label{eq:Lsigma}
\mathbf{L}(\Sigma)=\frac{\Ls^2}{2}\int_\Sigma\diff^2y\, \sqrt{\gamma}\, \mathcal K_{\Sigma }-2\pi \Ls^2\chi\left(\Sigma\right)\, .
\end{equation}
Namely, the LM prescription gives rise to a codimension-two functional that inherits the conformal invariance of the parent action. Furthermore, $\mathbf{L}$ is free of IR divergences for any boundary anchored surface $\Sigma$ embedded in an arbitrary bulk spacetime~\cite{Anastasiou:2024xxx}. Let us remember that until now, the embedding of the two-dimensional surface $\Sigma$ in a four-dimensional space is considered in complete generality, with the only requirement that $\Sigma$ is \textit{compact}.\footnote{This includes conformally compact surfaces in AdS, which have infinite area.} This is a constraint that we inherited from employing the Gauss-Bonnet theorem \eqref{eq:GBt}, and it will become relevant afterwards.
%Notice that conformal invariance and finiteness are features met by the finite part of the EE, $F(A)$ in the vacuum, due to its analogy to Willmore energy  $\mathbf W_3$~\eqref{eq:FAW}. 

As shown in Ref.~\cite{Anastasiou:2022ljq}, the energy functional $\mathbf{L}$ recovers in a certain limit not only  $\mathbf W_3$ but also the renormalized area $\mathbf{A}^{\text{ren}}$, which is related to the holographic EE universal term $F(A)$ for generic states. In what follows, we review the derivation of these energy functionals from $\mathbf{L}$, making manifest the significance of conformal symmetry in their construction. On top of that, a byproduct of the same functional is the reduced Hawking mass $\mathbf{I}_3$, which we also include. We present each case separately.

\subsection{Renormalized area}

In order to make explicit the relation between the renormalized area $\mathbf{A}^{\text{ren}}$ and the functional $\mathbf L$, it is particularly convenient to reexpress the invariant $\mathcal K_\Sigma$ in terms of the subtraces on $\Sigma$ of the bulk Weyl tensor $W_{ ab }^{ ab }$ and the square of the traceless part of the extrinsic curvature ${K^{A}}_{\langle ab \rangle}\equiv {K^{A}}_{ab}-\frac{1}{2}\gamma_{ab}K^{A}$ as
\begin{equation}\label{eq:KW}
    \mathcal K_{\Sigma }=W_{ ab }^{ ab }-{K^{A}}_{\langle ab \rangle}{K_{A}}^{\langle ab \rangle}\, .
\end{equation}
Since the area term in the RT formula~\eqref{RT} comes from the conical contribution of the Einstein-Hilbert action due to Eq.~\eqref{Rconical}, then, at the saddle point, the surfaces minimizing the area should be embedded into Einstein spacetimes. As a consequence, it is expected that renormalized area should result from the same class of ambient spacetimes but with AAdS asymptotics. Even though, the parent action of $\mathcal K_{\Sigma }$ is a higher-curvature gravity theory, \emph{i.e.}, CG, this admits an Einstein sector in its set of solutions~\cite{Anastasiou:2020mik,Anastasiou:2023oro}. For them, the Weyl tensor acquires the particularly simple expression of Eq.~\eqref{FAdS} for which we can exchange the Riemann tensor of the ambient space with quantities defined on the embedded surface using the Gauss-Codazzi relation
\begin{equation}
R_{ab}^{ab}=\mathcal R-K^{2} +{K^{A}}_{ab} {K_{A}}^{ab} \,.
\label{Riemsubtr}
\end{equation}
In turn, this implies that the $\mathbf L$ functional for Einstein spacetimes reads
\begin{equation}\label{eq:LSAren}
\mathbf{L}(\Sigma)\big|_{\text E}=\mathbf{A}^{\text{ren}}(\Sigma)-\frac{\Ls^2}{4}\int_\Sigma\diff^2y\sqrt{\gamma}\, K^{2}\, ,
\end{equation}
where we denoted
\begin{equation}
\mathbf{A}^{\text{ren}}(\Sigma)\equiv \frac{\Ls^2}{2}\int_\Sigma\diff^2y\sqrt{\gamma}\left(\mathcal R+\frac{2}{\Ls^2}\right)-2\pi \Ls^2\chi(\Sigma) \,,
\label{renArea}
\end{equation}
as the renormalized area of the two-dimensional embedded surface $\Sigma$ \cite{Fischetti:2016fbh,Anastasiou:2017xjr}. Notice that up to this point, the surface $\Sigma$ is not required to be minimal. However, it becomes manifest from the previous expression that $\mathbf{L}(\Sigma)\big|_{\text E}$ reduces to renormalized area for minimal surfaces.

The latter allows us to make contact with holographic EE. This is achieved by requiring that the submanifold of interest to be a RT surface $\Sigma=\RT$, \emph{i.e.}, cobordant and homologous to the entangling region $A$ under consideration with the additional requirement of being a minimal surface. The minimality condition is crucial for our expression \eqref{eq:LSAren} as it implies the vanishing of the trace of the extrinsic curvature $K^{A}=0$. As a consequence, we see that $\mathbf L$ reproduces the finite part of the EE \cite{Anastasiou:2022ljq}
\begin{equation}
F(A)=-\frac{\mathbf A^{\text{ren}}(\RT)}{4\GN}=-\frac{1}{4\GN}\mathbf{L}(\RT)\big|_{\text E}\, ,
\label{FAL}
\end{equation}
as long as we are considering an ambient Einstein spacetime and a RT surface.

\subsection{Reduced Hawking mass}

An interesting feature of Eq.~\eqref{eq:LSAren} is its applicability to a general class of surfaces, either minimal or non-minimal. Since for RT surfaces, which are minimal, one makes contact with the finite part of EE, it is necessary to understand its behavior when the minimality condition is lifted. In this case, the hypersurface $\Sigma$ remains unrestricted while being embedded in Einstein-AdS spacetimes, and Eq.~\eqref{eq:LSAren} can be cast as follows
\begin{equation}\label{eq:IfromL}
\mathbf{L}(\Sigma)\big|_{\text E}=\frac{\Ls^2}{4}\, \mathbf{I}_3(\Sigma)-2\pi\Ls^2\chi\left(\Sigma\right)\,,
\end{equation}
where
\begin{equation}
\mathbf{I}_3\left(\Sigma\right)=\int_\Sigma\diff^2y\sqrt{\gamma}\left[2\mathcal R+\frac{4}{\Ls^2}-K^2\right]\, .
\end{equation}
Here, $\mathbf I_3(\Sigma)$ is identified as the reduced Hawking mass, a generalization of the Hawking mass for AAdS spacetimes introduced in Ref.~\cite{Fischetti:2016fbh}. Namely, $\mathbf{L}(\Sigma)$ becomes the reduced Hawking mass, up to a topological contribution, when ambient Einstein-AdS spacetimes are considered. When the latter is evaluated on minimal surfaces $\Sigma$, one recovers the renormalized area functional. This object has very intriguing properties, since it is monotonous under inverse mean curvature flows. This feature allowed the authors of Ref.~\cite{Fischetti:2016fbh} to obtain bounds on the renormalized holographic EE for arbitrary regions on general states of three-dimensional CFTs. As a consequence, $\mathbf{L}(\Sigma)$ not only probes renormalized holographic EE but also imposes rather generic bounds that $F(A)$ has to satisfy. %In the vacuum, one recovers the bounds given in Eq.~\eqref{Fminim}. 
%\comment{move here:} that played a crucial role in the extension of the Penrose inequality to hyperbolic spaces~\cite{lee2013penroseinequalityasymptoticallylocally}. 

\subsection{Willmore energy}
Consider the conformal invariant $\mathcal K_\Sigma$, appearing in $\mathbf L(\Sigma)$, given in terms of the Weyl tensor \eqref{eq:KW}. We can decompose this contribution into a sum of the codimension-two subtraces of the Ricci and Schouten tensors as
\begin{equation}
W_{ab}^{ab}=R_{ab}^{ab}-2S_{a}^{a} \,.
\end{equation}
Taking into account the Gauss-Codazzi relation of Eq.~\eqref{Riemsubtr} we obtain
\begin{equation}
\mathcal K_\Sigma=\mathcal R-\frac{1}{2} K^{2}-2S_{a}^{a} \,.
\end{equation}
Until now, we always assumed $\Sigma$ to be compact. However, Willmore energy is a quantity defined for \textit{closed} surfaces ---\emph{i.e.}, compact surfaces without boundary. Because of this, we also assume that $\Sigma$ is closed for the time being. In turn, this means that we can invoke the two-dimensional Gauss-Bonnet theorem $\int_\cl\diff^2y\sqrt{\gamma}\, \mathcal R=4\pi\chi(\cl)$ to simplify the Euler characteristic with the Euler density, finding \cite{Anastasiou:2022ljq}
\begin{equation}
    \mathbf L(\cl)=-\frac{\Ls^2}{4}\int_{\cl}\diff^2y\sqrt{\gamma} \left(K^{2}+4S_{a}^{a}\right)\, .
\end{equation}
Interestingly, this expression is nothing less than the \textit{conformal Willmore energy} \cite{mondino2018global}, defined for a two-dimensional closed surface $\cl$ embedded in a Cauchy slice of a four-dimensional AAdS spacetime. Whenever the background space is pure AdS, we can relate it to the usual Willmore energy functional --- this is, for a closed surface embedded in $\mathbb R^3$. To see this, we perform a rescaling of the metric
\begin{equation}\label{eq:scaledg}
g_{\alpha\beta}=\text{e}^{2\varphi}\tilde g_{\alpha\beta}\, ,
\end{equation}
in which, in the case of a constant-time slice of Euclidean Poincaré-AdS space, $g_{\alpha\beta}\diff x^\alpha\diff x^\beta=\frac{\Ls^2}{z^2}\left(\diff z^2+\diff \mathbf{x}^2\right)$, with $\mathbf x=(x_1,x_2)$. We choose $\varphi=-\log \frac{z}{\Ls}$ to remove the conformal factor, arriving to the three-dimensional Euclidean space 
\begin{equation}
 \tilde g_{\alpha\beta}\diff x^\alpha\diff x^\beta=\diff z^2+\diff \mathbf{x}^2\, .
\end{equation}
%\emph{i.e.}, $\mathbb R^3$. 
Of course, this transformation also needs to be applied to the geometric functional we are considering. However, since $\mathbf L(\cl)$ is a conformally invariant quantity, we can replace immediately all terms by the rescaled ones. Since the Schouten tensor $\tilde S_{\alpha}^{\beta}$ vanishes identically in a flat background space, the rescaled functional $\mathbf L\left(\cl\hookrightarrow\mathbb{R}^3\right)$ reads \cite{Anastasiou:2022ljq}
\begin{equation}
    \mathbf L\left(\cl\hookrightarrow\mathbb{R}^3\right)=-\frac{\Ls^2}{4}\int_{\cl}\diff^2y\sqrt{\tilde \gamma} \, \tilde{K}^2=-\Ls^2\mathbf W_3\left(\cl\right)\, ,
\end{equation}
and reduces to the Willmore energy of the surface $\cl$ as given in Eq.~\eqref{eq:Wdef}, after expressing the mean curvature of $\cl$ in terms of its extrinsic curvature as $2\tilde{H}^A=\tilde{K}^A$.

In this derivation, we assumed that the surface under consideration is closed. Ultimately, we are interested in relating RT surfaces --- which are compact but not closed, as they possess a boundary --- to the Willmore energy. Because of this, we can resort to the prescription of doubling $\RT$ described in the beginning of the section. As a consequence, we have that 
\begin{equation}\label{eq:Lw}
\mathbf{L}\left(\RT\hookrightarrow\mathbb{R}^3\right)=-\frac{\Ls^2}{2}\mathbf W_3(2\RT)\, ,
\end{equation}
for an RT surface --- and, by extension, for any other compact surface $\Sigma$. This implies that, in holographic three-dimensional CFTs in the vacuum, the finite part of the EE can also be related to this writing of the functional $\mathbf L$, this is
\begin{equation}
F(A)=-\frac{1}{4\GN}\mathbf L\left(\RT\hookrightarrow\mathbb{R}^3\right)=\frac{\Ls^2}{8\GN}\mathbf W_3\left(2\RT\right)\, , 
\end{equation}
which is the expression presented in Eq.~\eqref{eq:FAW}. This relation has deep implications, since the global bounds characterizing the Willmore energy --- see Eq.~\eqref{eq:boundW} --- impose analogous constraints on $F(A)$ for holographic Einstein gravity --- see Eq.~\eqref{Fminim}.

As a direct consequence of the bound \eqref{eq:boundW} on the Willmore energy, we observe that, in line with \eqref{FAL}, we obtain a constraint on the renormalized area $\mathbf A^\text{ren}\left(\RT\right)$ of the RT surface \eqref{renArea}. Since there is a topological contribution --- through $\chi\left(\RT\right)$ ---, the Willmore energy bound leads to the condition \cite{Anastasiou:2020smm}
\begin{equation}
\int_\RT\diff^2y\sqrt{\gamma}\left(\mathcal R+\frac{2}{\Ls^2}\right) \leq 0 \,.
\end{equation}
This expression encodes information about the local properties of the minimal surface and is relevant to the study of its shape deformations.

\section{Holographic EE in $d=5$ as generalized Willmore energy} \label{sec:3}
In this section we follow the same line of reasoning as in the previous section, this time applied to the case of six-dimensional holographic Einstein gravity. In particular, we determine the four-dimensional conformally invariant functional which arises as the conical contribution of the six-dimenisonal CG with an Einstein sector. In analogy to its two-dimensional counterpart, this reduces to the renormalized holographic EE, giving rise to notions of renormalized area, reduced Hawking mass and, most importantly for our purposes, a \textit{generalized Willmore energy}. We show that this captures the universal contribution to the holographic EE for five-dimensional theories dual to Einstein gravity in the vacuum state. 

%The previous analysis indicates that energy functionals associated to the finite part of holographic EE for CFTs dual to Einstein gravity, demonstrate invariance under Weyl rescalings of the ambient metric. These are constructed out of CG that admits Einstein spacetimes in its solution space using the LM prescription. In this section we follow the same line of reasoning, aiming on constructing four dimensional generalizations of the same class of energy functionals. In particular, \textit{generalized Willmore energy} will allow us to establish several results concerning the shape-dependence of EE not just for holographic CFTs, but also for a more general class of theories at the boundary. To do so, we determine the four-dimensional conformally invariant functional $\mathbf F_\Sigma$ that arises as the conical contribution of the six-dimensional CG with an Einstein sector. In analogy to its two-dimensional counterpart, $\mathbf F_\Sigma$ will reduce to the renormalized holographic EE for generic CFTs through its relation to renormalized area and Willmore energy functionals. Furthermore, a proposal for the four-dimensional generalization of the reduced Hawking mass is presented.
%\comment{revisit later}

\subsection{CG in six dimensions}

Due to the fact that CG in six dimensions is a triparametric family of theories, seeking the combination with an Einstein subsector is a highly non-trivial task. However, conformal renormalization indicates that the renormalized Einstein-AdS action provides the seed that allows us to track down the desirable combination. Our starting point is the 
%Even though this novel scheme makes use of the extrinsic counterterms~\cite{Olea:2005gb}, the underlying principle is different. For this reason our starting point, will be the 
action of six-dimensional Einstein-AdS gravity enhanced by the third Chern form
\begin{equation}
   \mathcal B_5\equiv -6\deth\, \delta^{\mu_1\ldots\mu_5}_{\nu_1\ldots\nu_5}k^{\nu_1}_{\mu_1}\left(\frac{1}{4}r^{\nu_2\nu_3}_{\mu_2\mu_3}r^{\nu_4\nu_5}_{\mu_4\mu_5}-\frac{1}{3}r^{\nu_2\nu_3}_{\mu_2\mu_3}k^{\nu_4}_{\mu_4}k^{\nu_5}_{\mu_5}+\frac{1}{5}k^{\nu_2}_{\mu_2}k^{\nu_3}_{\mu_3}k^{\nu_4}_{\mu_4}k^{\nu_5}_{\mu_5}\right)\, , 
\end{equation}
namely,
\begin{equation}\label{eq:ItildeEren6}
\hat I_{\text{E}}^{\text{ren}}=\frac{1}{16\pi\GN}\left[\int_{\mathcal M}\diff^6 x\detg\, \left(R+\frac{20}{\Ls^2}\right)-\frac{\Ls^4}{72}\int_{\partial \mathcal M}\diff^5 X \mathcal B_5\right]\, .
\end{equation}
 The boundary term  can be conveniently rewritten as a bulk term, whose topological nature is made manifest by the  Euler theorem
\begin{align}\label{Eulertheorem}
   \int_{\mathcal{M}}\diff^6x\detg\, \mathcal{X}_{6} = 3!\left( 4\pi\right)^3\,\chi\left(\mathcal{M}\right) + \int_{\partial \mathcal{M}}\diff^5X\,\mathcal B_5\,,
\end{align}
where $\mathcal{X}_{6}=\frac{1}{8}\delta^{\alpha_1\ldots\alpha_{6}}_{\beta_1\ldots\beta_{6}}R^{\beta_1\beta_2}_{\alpha_1\alpha_2} \ldots R^{\beta_{5}\beta_{6}}_{\alpha_{5}\alpha_{6}}$ is the Euler density in six dimensions. Therefore, this form of the renormalized AdS action involves the cubic Lovelock term with a fixed coupling, namely,
\begin{equation} \label{Iren6Euler}
\tilde{I}_{\text{E}}^{\text{ren}}=\frac{1}{16\pi\GN}\int_\mathcal{M}\diff^6x\detg
\left( R+\frac{20}{\Ls ^{2}}-\frac{\Ls^{4}}{72}\mathcal{X}_{6}\right) + \frac{\pi^{2} \Ls^4}{3 \GN} \chi \left(\mathcal M\right) \,.
\end{equation}%
This combination renders the Einstein-Hilbert action finite for solutions whose boundary is conformally flat \cite{Anastasiou:2020zwc}. The tilde in the above functional makes reference to the fact that the finiteness is not achieved for an arbitrary AAdS geometry, but it is limited to the class just mentioned.
As a consequence, additional counterterms on top of the topological term $\mathcal{X}_{6}$ are required. At this point, it is difficult to think of an underlying principle which could give rise to such correction while also reproducing the topological term.

The proposal of Conformal Renormalization considers a symmetry enhancement at the level of the action: from general diffeomorphism  invariance to Weyl invariance \cite{Anastasiou:2020mik}. This feature becomes manifest by the vanishing of the local part of the action in Eq.~\eqref{Iren6Euler} for pure AdS spaces. As a consequence, the action is factorizable by the AdS curvature or, in other words, a given polynomial of the Weyl tensor for Einstein spaces --- a quantity that vanishes for AdS constant curvature configurations ---, \cite{Miskovic:2014zja,Anastasiou:2018mfk}
\begin{align}\label{EH2}
\tilde{I}_{\text{E}}^{\text{ren}} &= \frac{\Ls^4}{16\pi\GN}\int_{\mathcal{M}}\diff^6 x\detg\, P_6\left(W\big|_{\text E}\right)  + \frac{\pi^{2} \Ls^4}{3 \GN} \chi \left(\mathcal M\right)\,,
\end{align}
where the polynomial $P_6\left(W\big|_{\text E}\right)$ depends on 
\begin{equation}
    \mathcal Y_4\big|_{\text E}\equiv \delta_{\alpha_{1} \ldots \alpha_{4}}^{\beta_{1}\ldots \beta_{4}} W_{\beta_{1} \beta_{2}}^{\alpha_{1} \alpha_{2}}\Big|_{\text E}W_{\beta_{3} \beta_{4}}^{\alpha_{3} \alpha_{4}}\Big|_{\text E}\, , \quad \text{and} \quad  \mathcal Y_6\big|_{\text E}\equiv \delta_{\alpha_{1} \ldots \alpha_{6}}^{\beta_{1}\ldots \beta_{6}} W_{\beta_{1} \beta_{2}}^{\alpha_{1} \alpha_{2}}\Big|_{\text E}W_{\beta_{3} \beta_{4}}^{\alpha_{3} \alpha_{4}}\Big|_{\text E}W_{\beta_{5} \beta_{6}}^{\alpha_{5} \alpha_{6}}\Big|_{\text E}
\end{equation}
as
% contains quadratic and cubic contractions of $ W\big|_{\text E}$, 
% \begin{equation}
%  \mathcal Y_4\big|_{\text E}=\delta_{\alpha_{1} \ldots \alpha_{4}}^{\beta_{1}\ldots \beta_{4}} W_{\beta_{1} \beta_{2}}^{\alpha_{1} \alpha_{2}}\Big|_{\text E}W_{\beta_{3} \beta_{4}}^{\alpha_{3} \alpha_{4}}\Big|_{\text E}\, , \quad \text{and} \quad
%  \mathcal Y_6\big|_{\text E}=\delta_{\alpha_{1} \ldots \alpha_{6}}^{\beta_{1}\ldots \beta_{6}} W_{\beta_{1} \beta_{2}}^{\alpha_{1} \alpha_{2}}\Big|_{\text E}W_{\beta_{3} \beta_{4}}^{\alpha_{3} \alpha_{4}}\Big|_{\text E}W_{\beta_{5} \beta_{6}}^{\alpha_{5} \alpha_{6}}\Big|_{\text E}\, ,
% \end{equation}
%namely,
\begin{equation}
    P_6 \left(W\big|_{\text E}\right) =\frac{1}{2(4!)\Ls^2} \mathcal Y_4\Big|_{\text E}-\frac{1}{(4!)^2} \mathcal Y_6\Big|_{\text E}\,.
    \label{P6WE}
\end{equation}
This is a convenient rearrangement of the Einstein-AdS action with negative cosmological constant and, therefore, still a second-derivative theory. The presence of the Weyl tensor in the action is suggestive of the link to conformal symmetry. A suitable conformal covariantization of the above action would turn it into a particular form of CG in six dimensions. Such construction relies on a proper basis of six-derivative conformal invariants, given by \cite{Bonora:1985cq,Deser:1993yx,Erdmenger:1997gy,Bastianelli:2000rs,Oliva:2010zd,Metsaev:2010kp}
\begin{align}
I_1&\equiv W_{\alpha\beta\gamma\delta}W^{\alpha\lambda\eta\beta}W_{\lambda}{}^{\gamma\delta}{}_{\eta}\, , \label{I1}\\
I_2&\equiv W_{\alpha\beta\gamma\delta}W^{\gamma\delta\lambda\eta}W_{\lambda\eta}{}^{\mu\nu}\, ,\label{I2}\\
I_3&\equiv W_{\alpha\gamma\delta\lambda}\left(\delta_\beta^\alpha\dal+4R^\alpha_\beta-\frac{6}{5}\delta^\alpha_\beta R\right)W^{\beta\gamma\delta\lambda}+\nabla_\alpha J^\alpha\, ,\label{I3}
\end{align}
where the vector in the total derivative term is
\begin{equation}
J_\alpha\equiv R_\alpha{}^{\beta\gamma\delta}\nabla^\lambda R_{\lambda\beta\gamma\delta}+3R_{\beta\gamma\delta\lambda}\nabla_\alpha R^{\beta\gamma\delta\lambda}-R_{\beta\gamma}\nabla_\alpha R^{\beta\gamma}+\frac{1}{2}R\nabla_\alpha R-R_\alpha^\beta\nabla_\beta R+2R_{\beta\gamma}\nabla_\beta R^\gamma_\alpha \,.
\end{equation}
Naturally, the boundary term $\nabla_\alpha J^\alpha$ does not contribute to the equations of motion.

% The construction of a CG action as the conformal completion of the Einstein action supplemented with a topological term is a procedure far from rigorously defined. However, \textit{conformal covariantization} as a mechanism may be justified by the connection between every term in Eq.~\eqref{P6WE} and a given combination of the conformal invariants~\eqref{I1}-\eqref{I3}. 
The point now is to conformally covariantize action \eqref{EH2} into a CG action following the same procedure as the one discussed in Sec.~\ref{sec:21}. This time, we aim to bring it into a form involving the conformal invariants $I_1$, $I_2$, $I_3$. The most direct term to be conformally covariantized  is the one cubic in the Weyl tensor, that is,
\begin{equation}\label{WE3}
      \mathcal Y_6\Big|_{\text E} \xrightarrow{\text{c.c.}}  \mathcal Y_6 = 32(2I_{1}+I_{2})\,.
\end{equation}
On the other hand, the quadratic combination $\mathcal Y_4\big|_{\text E}$ in the polynomial cannot be directly related to conformal invariants in six dimensions. By itself, it can be cast as a six-derivative object by introducing the Schouten tensor. However, to restore Weyl covariance, we need to supplement it with a Cotton-squared  $C^2=C_{\alpha\beta\gamma}C^{\alpha\beta\gamma}$ (which vanishes for Einstein metrics) and a surface term $\hat{J}^\alpha=8 W^{\alpha\gamma\delta\beta}C_{\gamma\lambda\beta} - W^{\gamma\delta}_{\beta\varepsilon}\nabla^\alpha W^{\beta\varepsilon}_{\gamma\delta}$. Taking all this into account, we have\footnote{For more details on the conformal covariantization of action \eqref{EH2}, we refer the reader to Ref.~\cite{Anastasiou:2020mik}.}
\begin{equation}\label{eq:Y4cc}
    -\frac{1}{2\Ls^2} \mathcal Y_4\Big|_{\text E}\xrightarrow{\text{c.c.}}  I_4=\frac{1}{2}\delta_{\alpha_1\ldots\alpha_5}^{\beta_1\ldots\beta_5}W^{\alpha_1\alpha_2}_{\beta_1\beta_2}W^{\alpha_3\alpha_4}_{\beta_3\beta_4}S^{\alpha_5}_{\beta_5} + 8C^2+\nabla_\alpha \hat{J}^\alpha=  \frac{1}{3}\left(4I_{1}-I_{2}-I_{3}\right) \,.
\end{equation}
The explicit steps for this conformal covariantization are presented in Appendix~\ref{confcomple}. While in higher even dimensions a number of Schouten tensors may be inserted into the gravity action, which when evaluated on Einstein spaces become proportional to Kronecker deltas,  $S^\alpha_{\beta}\big|_{\text E}=-\frac{1}{2\Ls^2}\delta^\alpha_{\beta}$, there may be plenty of higher-derivative terms which are identically zero in the Einstein sector of the gravity theory. Fortunately, the six-dimensional case is simple enough for Weyl invariance to remove the ambiguities in the couplings of the different terms in the CG action.

Putting together Eqs.~\eqref{WE3} and \eqref{eq:Y4cc}, we see that by conformally covariantizing the polynomial $P_{6}$ as given in Eq.~\eqref{P6WE}, one ends up with the CG that admits an Einstein sector introduced by Lü, Pang and Pope in Ref.~\cite{Lu:2013hx}, namely
\begin{equation}\label{eq:LPP1}
 -(4!)P_6 \left(W\big|_{\text E}\right) \xrightarrow{\text{c.c}}\mathcal C=4 I_{1} + I_{2} -\frac{1}{3} I_{3}\, .
\end{equation}
%One ends up, with the CG that admits an Einstein sector introduced by Lu, Pang and Pope in Ref.~\cite{Lu:2013hx}. 
The corresponding Lagrangian density with the associated surface term is given by\footnote{As a matter of fact, it was proven in Ref.~\cite{Lu:2013hx} that the Schwarzschild-AdS black hole is a solution of the higher-derivative equations of motion of this gravity theory. A more compact expression for the field equations --- in terms of the Weyl, Cotton and Schouten tensors --- obtained in Refs.~\cite{Anastasiou:2020mik} and~\cite{Anastasiou:2023oro} readily implies that  Einstein spaces constitute a proper sector of LPP CG, in a similar fashion to the four-dimensional case.}
\begin{equation}\label{ICG2}
    I_{\rm LPP} = \alpha\int_{\mathcal{M}}\diff^6 x\sqrt{|g|}\,\mathcal{L}_{\rm LPP}-2(4\pi)^3\alpha\chi \left(\mathcal M\right)+ \alpha\int_{\partial \mathcal{M}} \diff^5 x\sqrt{|h|}\,n_\alpha \hat{J}^\alpha \,,
\end{equation}
where
\begin{equation}
    \mathcal{L}_{\rm LPP}=\frac{1}{4!}\mathcal Y_6 + \frac{1}{2}\delta_{\alpha_1\ldots\alpha_5}^{\beta_1\ldots\beta_5}W^{\alpha_1\alpha_2}_{\beta_1\beta_2}W^{\alpha_3\alpha_4}_{\beta_3\beta_4}S^{\alpha_5}_{\beta_5} + 8C^2\, ,
\end{equation}
and where we added a topological contribution to make contact with Eq.~\eqref{EH2}. This action admits Einstein spacetimes in its set of solutions as shown explicitly in Ref.~\cite{Anastasiou:2023oro}.

% \begin{align}\label{LLPP}
%     \mathcal{L}_{\rm LPP} &= \alpha\left(\frac{1}{4!}\mathcal Y_6 + \frac{1}{2}\delta_{\alpha_1\ldots\alpha_5}^{\beta_1\ldots\beta_5}W^{\alpha_1\alpha_2}_{\beta_1\beta_2}W^{\alpha_3\alpha_4}_{\beta_3\beta_4}S^{\alpha_5}_{\beta_5} + 8C^{\alpha\beta\gamma}C_{\alpha\beta\gamma}\right)\,, \\
%    \label{Jmu}
%     \hat{J}^\alpha &= \alpha\left(8 W^{\alpha\gamma\delta\beta}C_{\gamma\lambda\beta} - W^{\gamma\delta}_{\beta\varepsilon}\nabla^\alpha W^{\beta\varepsilon}_{\gamma\delta}\right)\,.
% \end{align}
Remarkably, evaluating the action of LPP CG for Einstein spacetimes, it reduces to the topologically renormalized action of Eq.~\eqref{EH2} enhanced by a total derivative contribution, up to the Euler characteristic
\begin{equation}\label{LPPEinstein}
I_{\rm{LPP}}\big|_{\text E} =-4!\alpha{\int_{\mathcal{M}}}\diff^{6}x\detg\,
P_{6}\left( W\big|_{\text E}\right)-2(4\pi)^3\alpha\chi \left(\mathcal M\right)-\frac{\alpha}{2} \int_{\partial \mathcal{M}}\diff^{5}X\sqrt{|h|}\,n^{\alpha }J_{\alpha}\big|_{\text E} \,,
\end{equation}%
where $J_{\alpha}\big|_{\text E} \equiv \frac{1}{2}\nabla _{\alpha} \left(W^{\beta \gamma}_{\delta \kappa}\big|_{\text E}  W^{\delta \kappa}_{\beta \gamma}\big|_{\text E}\right)$. The connection between CG  --- in the particular form of LPP action --- and the Einstein sector of the theory is made manifest by a suitable choice of the  coupling, $\alpha=-\Ls ^{4}/(384\pi \GN)$. Interestingly, the resulting boundary term renders the action of Eq.~\eqref{LPPEinstein} finite for Einstein-AdS spacetimes with a generic boundary geometry, namely, it recovers the renormalized Einstein action~\cite{Anastasiou:2020mik}.

When the asymptotic behavior of the spacetime is taken into account, in the form of a Fefferman-Graham expansion, the extra boundary term produces a new counterterm which is quadratic in the Weyl tensor of the boundary metric
%\begin{align}
%I_{\rm{LPP}}\left[ E\right] &=\frac{1}{16\pi \GN}{\int_{\mathcal{M}}}\diff^{6}x\detg\left( R+\frac{20}{\Ls ^{2}}-\frac{\Ls ^{4}}{72}\mathcal{X}_{6}\right) +\frac{\Ls ^{3}}{192\pi \GN}\int_{\partial \mathcal{M}}d^{5}X\deth\,
%\mathcal{W}^{\mu\nu \sigma \tau}(h)\mathcal{W}_{\mu \nu \sigma \tau}(h)\,, \notag\\
%&= I_{\text{E}}^{\text{ren}}\,.
%\end{align}
\begin{equation}
\small I_{\rm{LPP}}\big|_{\text E} =\frac{1}{16\pi \GN}\left[\int_{\mathcal{M}}\diff^{6}x\detg\left( R+\frac{20}{\Ls ^{2}}-\frac{\Ls ^{4}}{72}\mathcal{X}_{6}\right)+\frac{16}{3}\pi^3\Ls^4\chi \left(\mathcal M\right) +\frac{\Ls ^{3}}{12}\int_{\partial \mathcal{M}}\diff^{5}X\deth\, w^2\right]= I_{\text{E}}^{\text{ren}}\,,
\end{equation}
plus other terms which vanish as $\partial \mathcal{M}$ is taken to the conformal boundary. Here, we denoted by $w^2=w_{\mu\nu}^{\rho\sigma}w^{\mu\nu}_{\rho\sigma}$ the Weyl-squared tensor at the AdS boundary. This counterterm correctly removes divergences induced by nontrivial conformal properties of the boundary metric. This result indicates the profound relation between bulk Weyl symmetry and renormalized Einstein-AdS action.

\subsection{Energy functionals coming from LPP CG}
Following the idea of the four-dimensional case, we construct the codimension-two functionals which are invariant under Weyl rescalings of the ambient spacetime, starting from the unique combination of six-dimensional CG that admits Einstein-AdS spacetimes in its solution space. As discussed previously, this is achieved by a precise combination of the three conformal invariants $\mathcal C$, given in Eq.~\eqref{eq:LPP1}.
%~\cite{Lu:2011ks} 
%\begin{equation}\label{eq:LPP1}
%\mathcal C=4 I_{1} + I_{2} -\frac{1}{3} %I_{3} \,.
%\end{equation}
In particular, our starting point will be the action \cite{Lu:2011ks,Anastasiou:2020mik}
\begin{equation}\label{eq:LPP2}
I_{\text{LPP}}=-\frac{\Ls^4}{384 \pi \GN} \int_{\mathcal M} \diff^{6}x \sqrt{|g|} \,\mathcal C+ \frac{\pi^{2} \Ls^4}{3 \GN} \chi \left(\mathcal M\right) \,.
\end{equation}
Extending the prescription of the previous section to six dimensions, we evaluate the invariants $I_1$, $I_2$ and $I_3$ in the orbifold. However, the situation is more delicate than in four dimensions. In general, cubic curvature invariants are sensitive to the so-called \textit{splitting problem} \cite{Miao:2014nxa,Camps:2014voa,Miao:2015iba}. This means that there are different naive ways to regularize the action near the conical singularity and, thus, different holographic EE functionals are obtained depending on the regularization scheme.\footnote{There are higher-curvature theories for which this issue does not arise, such as quadratic gravity \cite{Fursaev:2013fta, Anastasiou:2021swo}, $f(R)$ gravity or Lovelock gravity \cite{Camps:2013zua,Hung:2011xb,deBoer:2011wk,Anastasiou:2021jcv}. This also occurs when the coupling constants of any higher-curvature term are treated perturbatively \cite{Bueno:2020uxs}.} However, it was shown in Ref.~\cite{Miao:2015iba} that bulk conformal symmetry should be induced to the resulting codimension-two functional. This allows to parametrize a family of splittings where the universal part of the holographic EE is independent of the specific choice. Exploiting this remarkable property of the combination \eqref{eq:LPP1}, one ends up with the following expression \cite{Miao:2015iba}
\begin{equation}
    \mathcal C\left(\mathcal M^{(\vartheta)}\right)=\mathcal C\left(\mathcal M\right)+2\pi (1-\vartheta)\mathcal C_\Sigma\, ,\quad \mathcal C_\Sigma=4 F_{1} + F_{2} -\frac{1}{3} F_{3}\, ,
\end{equation}
where the codimension-two invariants $F_i$ for $i=1,2,3$ come from each of the $I_i$ and can be cast in the form
\begin{align}
F_{1} =& 3 \left(W^{\alpha \beta \gamma \delta }W_{\beta \ \ \gamma }^{\ \lambda \eta }\varepsilon _{\lambda \alpha }\varepsilon _{\delta \eta } -\frac{1}{4}W^{\lambda \delta \eta \beta }W_{\ \delta \eta \beta }^{\alpha }g_{\alpha \lambda }^{\perp } +\frac{1}{20}W^{\alpha\beta\gamma\delta}W_{\alpha\beta\gamma\delta} \right)+3 K^{\iota }{}_{\langle \lambda \alpha\rangle} K_{\iota}{}^{\langle\beta \gamma\rangle}W^\lambda{}_\beta{}^\alpha{}_\gamma  \nonumber \\
&-3 K^{\iota}{}_{\ \langle\lambda \alpha\rangle}K_{\iota}{}_{\langle\beta \gamma\rangle} K_{\zeta }{}^{\langle\lambda [\gamma\rangle}K^{\zeta}{}^{\langle\alpha]\beta\rangle}+3\varepsilon ^{\iota \zeta } K_{\iota\langle\lambda \eta\rangle}K_{\zeta}{}^{\langle\alpha\eta \rangle}\varepsilon^{\gamma \delta }W^\lambda{}_{\alpha\gamma \delta} +\frac{3}{4}\left (K^{\iota }{}_{\langle\alpha \beta\rangle} K_{\iota }{}^{\langle\alpha \beta \rangle}\right )^{2} \nonumber \\
&+3\varepsilon ^{\iota \zeta}\varepsilon ^{\kappa \delta }K_{\iota\langle\lambda \eta\rangle}K_{\zeta}{}^{\langle\alpha \eta\rangle} K_{\kappa}{}^{\langle\gamma\lambda\rangle}K_{\delta}{}_{\langle\gamma \alpha \rangle}-\frac{3}{4} K^{\iota }{}_{\langle\lambda \eta\rangle} K_{\iota }{}^{\langle\lambda \eta\rangle}W^{\alpha \beta \gamma \delta }\varepsilon _{\alpha \beta }\varepsilon _{\gamma \delta }\,,\\ \notag 
F_{2} =&3 \left(W^{\alpha \beta \gamma \delta }W_{\gamma \delta }^{\lambda \eta }\varepsilon _{\lambda \eta }\varepsilon _{\alpha\beta} -W^{\lambda \delta \eta \beta } W^{\alpha}{}_{\delta \eta \beta }g_{\alpha \lambda }^{\perp } +\frac{1}{5}W^{\alpha\beta\gamma\delta}W_{\alpha\beta\gamma\delta}\right)\\
&-6 K^{\iota }{}_{\langle\delta}{}^{(\gamma\rangle}K_{\iota }{}^{\langle\alpha )\delta\rangle}\left(2W_{\beta \gamma \lambda \alpha }g^{\perp \beta \lambda }+K^{\zeta }{}_{\langle\alpha \eta\rangle}K_{\zeta\langle\gamma}{}^{\eta\rangle}\right)\\&+6\varepsilon ^{\alpha\beta}\varepsilon ^{\gamma\delta}K_{\alpha}{}^{\langle\zeta\iota\rangle}K_{\beta}{}^{\langle\eta}{}_{\iota\rangle}\left(2W_{\gamma\eta\delta\zeta}+K_{\delta\langle\eta \lambda\rangle}K_{\delta}{}^{\langle\lambda}{}_{\zeta\rangle}\right)\,, \nonumber \\
F_{3} =& -6\mathcal X_4^\Sigma+12F_{1} +3F_{2} +192 \left(\Upsilon _{a}^{a} -\frac{1}{2} S_{ab}S^{ab} +\frac{1}{4}\left (S_{a}^{a}\right )^{2} -\frac{1}{4} K^{A} K_{A cb}S^{cb}  \right. \nonumber\\
& \left. +\frac{3}{32}K^{A}K_{A } S_{b}^{b} -\frac{1}{16} K^{A }K^{B } S_{AB } -\frac{1}{32}K^{A }K_{A cb}K_{B }K^{B cb} +\frac{7}{1024}\left (K^{A }K_{A }\right )^{2}\right) \,,
\label{Ficonical}
\end{align}
where
\begin{align}
\Upsilon_{a b} &=\frac{1}{4}\left[\frac{1}{16}\left(\partial _{a}K^{A} \partial _{b}K_{A} +K_{A}K^{A}{}_{ac}K_{B}K^{B}{}_b^c -K^{A}K^{B} W_{aAbB}-K^{A}K_{A}S_{a b} \right.\right.\nonumber  \\
&\left.-K^{A}K^{B}S_{AB}\gamma_{a b}\right) +S_{a\alpha}S_{b}^{\alpha} -B_{a b}-\frac{1}{2}\left(S_{aA} \partial _{b}K^{A} -S_a^cK^{A}{}_{bc}K_{A} +K^{A}C_{a bA}\right. \nonumber  \\
&\left.\left.+\nabla_{a}^{\Sigma}\left (K^{A}S_{Ab}\right ) -K^{A}K^{B}{}_{ab}S_{AB} \right)\right]\,,
\label{upsilonfin}
\end{align}
and $K^{\gamma}{}_{\alpha\beta}$ is defined in Eq.~\eqref{eq:covK}. For $F_1$ and $F_2$, we chose to maintain the covariant formulation in order to simplify the resulting expressions, considering $\varepsilon _{\alpha\beta} = n_{\alpha}^{A} n_{\beta}^{B} \epsilon_{AB}$ as the binormal, where $\epsilon_{AB}$ is the Levi-Civita tensor, and identifying $g^{\perp}_{\alpha\beta}$ as the two-dimensional metric of the normal bundle. For later convenience, we perform the normal decomposition of the $F_3$, by labeling the bulk indices as $\alpha=\left(A,a\right)$, where $a$ denotes tangential indices and $A$, the normal directions to $\Sigma$. Here, $C_{\alpha\beta\gamma}$ and  $B_{\alpha\beta}$ represent the Cotton and Bach tensor of $\mathcal{M}$, respectively, where $\nabla_a^{\Sigma}$ is the covariant derivative with respect to the codimension-two intrinsic metric $\gamma_{ab}$. Further details on the normal decomposition of $F_3$ are given in Appendix~\ref{Appendixcomputation}.

As a consequence, the evaluation of LPP CG in six dimensions in the conical singular manifold gives rise to a conformal codimension-two functional
\begin{equation}
I_{\text{LPP}} \left (\mathcal M^{(\vartheta)}\right )= I_{\text{LPP}} +\frac{\left(1-\vartheta\right)}{4\GN}\, \mathbf{F}(\Sigma)\, ,   
\end{equation}
where $\mathbf F(\Sigma)$ denotes the conical part whose explicit expression reads
\begin{equation}\label{eq:FS}
\mathbf F(\Sigma)=-\frac{\Ls^4}{48}\int\diff^4y\sqrt{\gamma}\, \mathcal C_\Sigma+\frac{4}{3}\pi^2\Ls^4\chi(\Sigma)\,.
\end{equation}
Here, we took advantage of the self-replicating property of the Euler-characteristic in its codimension-two version when evaluating it in the orbifold $\chi\left(\mathcal M^{(\theta)}\right)=\chi\left(\mathcal M\right)+\left(1-\vartheta\right)\chi\left(\Sigma\right) $.
%\begin{equation}
%\chi\left(\mathcal M^{(\theta)}\right)=\chi\left(\mathcal M\right)+\left(1-\vartheta\right)\chi\left(\Sigma\right) \,.
%\end{equation}
Interestingly, the functional $\mathbf F(\Sigma)$ is the four-dimensional extension of the Graham-Witten anomaly $\mathbf L(\Sigma)$, dubbed Graham-Reichert anomaly.\footnote{In Ref.~\cite{Graham:2017bew} it was shown that the $\mathbf F(\Sigma)$ functional appears as the logarithmic coefficient in the asymptotic expansion of the area for codimension-two minimal boundary-anchored surfaces embedded in seven-dimensional asymptotically hyperbolic Einstein manifolds, which is the area anomaly by definition.}

After some algebraic manipulation, we rewrite $\mathbf F(\Sigma)$ in a simplified form as
\begin{align}
\mathbf{F}(\Sigma)&=-\frac{4\Ls^4}{3}\int _{\Sigma }\diff^4y\sqrt{\gamma}\left[\frac{1}{32}\mathcal X_4^\Sigma+\Upsilon_a^a +\frac{1}{2} S_{ab}S^{ab} -\frac{1}{4}\left (S_a^a\right )^{2} +\frac{1}{4} K^A K_{A ab}S^{ab} -\frac{3}{32}K^A K_A S_a^a\right.\nonumber  \\
&\left.+\frac{1}{16} K^A K^B S_{AB}+\frac{1}{32}K^A K_{A ab}K_B K^{B ab}-\frac{7}{1024}\left (K^A K_A \right )^2\right]+\frac{4\pi ^{2}\Ls^4}{3}\chi \left (\Sigma \right ) +\text{b.\,t.}\, ,\label{eq:FS2}
\end{align}
where $\mathcal X_4^\Sigma=\frac{1}{4}\delta_{abcd}^{efgh}\mathcal R_{ef}^{ab}\mathcal R_{gh}^{cd}$ is the Euler-density of the four-dimensional surface $\Sigma$ and the boundary terms (b.\,t.) come from the term $\nabla_\alpha J^\alpha$ in $I_3$ --- these terms are dropped in Ref.~\cite{Miao:2015iba}, however they render $F_3$ a conformal invariant for a manifold with boundaries.

This exact form will allow us to make contact with higher-dimensional analogues of both the reduced Hawking mass and Willmore energy.  However, the computation of the former strongly depends on the determination of the boundary term in Eq.~\eqref{eq:FS2}, which is a quite challenging task due to he presence of covariant derivatives, what makes the computation of their conical part rather complicated. Interestingly, the form of the boundary term will be greatly simplified when restricting ourselves to Einstein spacetimes. The four-dimensional analysis in Sec.~\ref{sec:2} indicates that both functionals arise when considering surfaces embedded in this exact class of spacetimes. 

Based on these considerations, in the next subsection we proceed in our quest of determining the 
codimension-two functionals. For a better presentation of the argument, we find it is more convenient to discuss first the reduced Hawking mass and renormalized area functionals, and then end with the generalized Willmore energy and its connection to $F(A)$ in $d=5$.

\subsection{Generalized reduced Hawking mass and renormalized area}

Obtaining a candidate for the reduced Hawking mass of a four-dimensional hypersurface embedded in a six-dimensional bulk spacetime is a highly non-trivial task, since this has to meet the same criteria as its two-dimensional analogue, but for a functional constructed out of four-derivative objects. As it was revealed in Ref.~\cite{Anastasiou:2022ljq}, and discussed in the previous section, the two-dimensional reduced Hawking mass arises from the Graham-Witten anomaly functional $\mathbf{L}(\Sigma)$ when evaluated for Einstein spacetimes~\eqref{eq:IfromL}. Similarly, we present a candidate for the four-dimensional reduced Hawking mass, as the functional coming from the evaluation of the Graham-Reichert anomaly $\mathbf F(\Sigma)$ in Einstein spacetimes. In this case, the missing boundary contribution in Eq.~\eqref{eq:FS2} --- which we denoted $\text{b.\,t.}$ --- can be determined, as it comes entirely from the $I_{3}$ conformal invariant. Indeed, the surviving boundary contribution from the CG action \eqref{eq:LPP2} when evaluated in the Einstein sector is
\begin{equation}\label{eq:J}
J\big|_{\text E}=\frac{\Ls^{4}}{384 \pi \GN}\int _{ \partial \mathcal M}\diff^{5}X\deth\, n^{\alpha}J_{\alpha}\big|_{\text E}=-\frac{\Ls^{3}}{192 \pi \GN}\int _{ \partial \mathcal M}\diff^{5}X\deth\, w^2 \,,
\end{equation}
where $w^2=w_{\mu\nu}^{\rho\sigma}w^{\mu\nu}_{\rho\sigma}$ is the Weyl-squared tensor at the AdS boundary. The last equality is valid only asymptotically, but this is sufficient since its purpose is just to cancel bulk divergences~\cite{Anastasiou:2020mik}.

Now, we have to evaluate Eq.~\eqref{eq:J} in the conically singular manifold and find the codimension-two contributions. Since we already encountered the decomposition of the Weyl-squared tensor in Eq.~\eqref{eq:W2c}, we just need to adapt the decomposition to the submanifold $\partial \mathcal M$ in order to find the boundary terms anticipated in Eq.~\eqref{eq:FS2}, this is
\begin{equation}\label{eq:btE}
\text{b.\,t.}\big|_{\text E}= -\frac{\Ls^{3}}{6}\int_{ \partial \Sigma }\diff^{3}Y\sqrt{\sigma}\left (w_{ij}^{ij} -\kappa^{I}{}_{\langle ij\rangle}\kappa_{I}{}^{\langle ij\rangle}\right )=-\frac{\Ls^3}{6}\int_{\partial\Sigma}\diff^3Y\sqrt{\sigma}\, \mathcal K_{\partial\Sigma}\,.
\end{equation}
Here, $\kappa^{I}{}_{\langle ij\rangle}$ is the traceless extrinsic curvature of $\partial \Sigma$ embedded in $\partial \mathcal M$. In the last equality, we make manifest that the boundary term evaluated in Einstein spacetimes reproduces the functional obtained in Eq.~\eqref{eq:KW}, this time for boundary manifolds.

Finally, we are ready to combine all our partial results regarding the evaluation of the functional $\mathbf F(\Sigma)$ in Einstein spacetimes. In this case, both the Bach and the Cotton tensors vanish identically and the Schouten turns proportional to the metric. As a consequence, the resulting expression is simplified a lot. In particular, starting from the Graham-Reichert formula of Eq.~\eqref{eq:FS2} along with $\Upsilon_a^a$ as given in Eq.~\eqref{upsilonfin} and the boundary term \eqref{eq:btE}, we obtain a functional $\mathbf F(\Sigma)\big|_{\text E}$ from which we identify a generalized reduced Hawking mass $\mathbf I_5(\Sigma)$ defined for four-dimensional submanifolds $\Sigma$,
\begin{equation}\label{eq:IfromF}
\mathbf{F}(\Sigma)\big|_{\text E}=\Ls^4\mathbf I_5(\Sigma)+\frac{4\pi^3\Ls^4}{3}\chi\left(\Sigma\right)\, .
\end{equation}  
The explicit expression is
\begin{align}
\mathbf I_5(\Sigma)=&\frac{1}{48}\int _{\Sigma}\diff^4y\sqrt{\gamma}\left[\frac{48}{\Ls^{4}} -\mathcal X_4^{\Sigma} + (\partial K)^2 -K K^{ab} K K_{ab}+\frac{7}{16}K^4-\frac{6}{\Ls^{2}}K^2\right.  \nonumber  \\
&-K_{A}K^{B}\left(\mathcal R_{iB}^{i A}+\frac{1}{\Ls^2}\delta_{i B}^{iA}\right) \bigg]-\frac{1}{6\Ls}\int_{\partial\Sigma}\diff^3Y\sqrt{\sigma}\, \mathcal K_{\partial\Sigma}\big|_{\text E}\, ,\label{eq:RHM5}
\end{align}
where $(\partial K)^2\equiv \partial ^{a}K^{A} \partial _{a}K_{A}$, $K K^{ab} K K_{ab}\equiv K^{A}K_{A}{}^{ab}K_{B}K^{B}{}_{ab}$ and $K^{4}\equiv \left (K^{A}K_{A}\right )^{2}$. This is one of our new results so let us make some observations. $\mathbf I_5\left(\Sigma\right)$ is a conformal invariant for any codimension-two surface $\Sigma$ embedded in an Einstein spacetime. Following its two-dimensional counterpart, it is expected to be free of infrarred (IR) divergences for any boundary-anchored surface, either extremal or non-extremal. Furthermore, the presence of the Gauss-Bonnet density is a desirable feature regarding the monotonous evolution of the functional along inverse mean curvature flows.\footnote{We thank S. Fischetti for the comments.}

A particularly interesting case arises when $\Sigma$ is an extremal surface, $\Sigma_{\text{ext}}$. In this situation, the vanishing of the trace of the extrinsic curvature $K^{A}=0$ simplifies significantly the form of $\mathbf{F}(\Sigma)$, which now reduces to the renormalized area of $\Sigma$, $\mathbf{F}(\Sigma_{\text{ext}})\big|_{\text E}=\mathbf A^{\text{ren}}(\Sigma_{\text{ext}})$, where
\begin{equation}\label{eq:renA4}
\mathbf A^{\text{ren}}(\ext)=\mathbf A(\ext)-\frac{\Ls^4}{24}\int_{\ext}\diff^4y\sqrt{\gamma}\,\mathcal X_4^{\Sigma}+\frac{4\pi^2\Ls^4}{3}\chi(\Sigma)-\frac{\Ls^3}{6}\int_{\partial\ext}\diff^3Y\sqrt{\sigma}\, \mathcal K_{\partial\ext}\,.
\end{equation}
Since we are interested in holographic EE, we will use this result for RT surfaces, which are a subclass of extremal surfaces.

Finally, let us recall that in the derivation of the renormalized area $\mathbf A^{\text{ren}}(\Sigma_{\text{ext}})$ we made use of the Gauss-Bonnet theorem \eqref{eq:GBt}, which requires $\Sigma_{\text{ext}}$ to be a compact surface. In order to find an expression for extremal non-compact surfaces, we must undo the exchange of the Euler density and characteristic of $\Sigma_{\text{ext}}$, this means
\begin{equation}\label{eq:renA4nc}
\mathbf A^{\text{ren}}(\Sigma_{\text{ext}})=\mathbf A(\ext)-\frac{\Ls^3}{6}\int_{\partial \ext}\diff^3 Y\left[\frac{\Ls}{4}\mathcal B_3^{\ext}+\sqrt{\sigma}\mathcal K_{\partial \ext}\right]\,,
\end{equation}
where $\mathcal B_3^{\Sigma_{\text{ext}}}=-\frac{2}{3}\sqrt{\sigma}\delta^{lmn}_{ijk}\mathfrak K^i_l\left(3\mathfrak R^{jk}_{mn}-2\mathfrak K^j_m\mathfrak K^k_n\right)$ is the explicit expression of the second Chern form associated to $\Sigma_{\text{ext}}$.

By particularizing the extremal surface to be an RT surface $\RT$, it is immediate to find the finite piece of holographic EE in $d=5$ from the renormalized area \eqref{eq:renA4nc} of $\RT$ as
\begin{equation}\label{FAL5}
F(A)=\frac{\mathbf A^{\text{ren}}(\RT)}{4\GN}=\frac{1}{4\GN}\mathbf{F}(\RT)\big|_{\text E}\, .
\end{equation}

\subsection{Generalized Willmore energy}\label{sec:GW4}

Once we have an expression for the functional $\mathbf F(\Sigma)$ at our disposal in Eq.~\eqref{eq:IfromF}, it is straightforward to construct a quantity that corresponds to the generalized Willmore energy. Following the same reasoning as in Sec.~\ref{sec:21}, we will initially assume that we are dealing with a closed surface $\cl$. In a later stage we will deal with non-closed ones --- like an RT surface. In this case the boundary terms can be dropped, leading to
\begin{align}
    \mathbf{F}(\cl)=&-\frac{4\Ls^4}{3}\int _{\Sigma_\text{cl}}\diff^4y\sqrt{\gamma}\left[\Upsilon_a^a +\frac{1}{2} S_{ab}S^{ab} -\frac{1}{4}\left (S_a^a\right )^{2} +\frac{1}{4} K^A K_{A ab}S^{ab} -\frac{3}{32}K^A K_A S_a^a\right.\nonumber  \\
&\left.+\frac{1}{16} K^A K^B S_{A B}+\frac{1}{32}K^A K_{A ab}K_B K^{B ab}-\frac{7}{1024}\left (K^A K_A\right )^2\right]\, .\label{eq:FS2c}
\end{align}
Here, we considered the Gauss-Bonnet theorem \eqref{eq:GBt} for the four-dimensional closed surface, 
\emph{i.e.}, $\int_\cl\diff^4 x\sqrt{\gamma}\, \mathcal X_4^\cl=32\pi^2\chi\left(\cl\right)$ in order to cancel the Euler density and characteristic. The formal definition of Willmore energy requires the embedding of $\Sigma$ into $\mathbb {R}^5$. This is achieved by starting with a constant time slice of a Euclidean global AdS$_{6}$ bulk spacetime and then choosing a convenient rescaling, like in Eq.~\eqref{eq:scaledg}, to map it to $\mathbb{R}^5$. The functional form of $\mathbf{F}(\cl)$ will not be modified, since it is conformally invariant, however this will be further simplified since all curvatures vanish when a flat ambient spacetime is considered. As a consequence, we end up in 
\begin{equation}\label{eq:GW5}
    \mathbf{F}\left(\cl\hookrightarrow\mathbb R^5\right)=\Ls^4\int_{\cl}\diff^4y\sqrt{\tilde\gamma}\, \mathcal{J}_{\cl}=\Ls^4\mathbf W_5\left(\cl\right) \,,
\end{equation}
where
\begin{equation}
    \mathcal{J}_{\cl}=\frac{1}{48}\left[(\partial\tilde K)^2-\tilde K \tilde K^{ab}\tilde K\tilde K_{ab}+\frac{7}{16}\tilde K^4\right]\, ,\label{eq:J5}
\end{equation}
where the tildes indicate that the quantities are evaluated in $\mathbb R^5$ as ambient space. This expression coincides with the generalization of the Willmore energy for closed four-dimensional surfaces given previously in the mathematical literature\footnote{Note that in the derivation presented in Ref.~\cite{guven2005conformally}, a factor of $-2$ is dropped in the
calculations, obtaining in the end incorrect coefficients for $\partial_a\tilde K^{A}\partial^a \tilde K_{A}$ and $\tilde K^{A}\tilde K_{A}{}^{ab}\tilde K_{B}\tilde K^{B}{}_{ab}$.} \cite{guven2005conformally,Gover:2016buc,zhang2017grahamwittens,Graham:2017bew,Blitz:2021qbp}.

For the case of a four-dimensional RT surface $\RT$, which is a compact but not closed submanifold, we apply the doubling of the $\RT$ prescription, described in Sec.~\ref{sec:21}. After doing so, we obtain the expression
\begin{equation}
\mathbf{F}\left(\RT\hookrightarrow\mathbb{R}^5\right)=\frac{\Ls^4}{2}\mathbf W_5\left(2\RT\right)\, ,
\end{equation}
which matches Eq.~\eqref{eq:Lw} but in two-dimensions higher. In analogy to the case of two-dimensional RT surfaces, we can make manifest the relation between the finite part of six-dimensional holographic EE and generalized Willmore energy as
\begin{equation}\label{eq:FG}
F(A)=\frac{1}{4\GN}\mathbf{F}\left(\RT\hookrightarrow\mathbb{R}^5\right)=\frac{\Ls^4}{8\GN}\mathbf W_5\left(2\RT\right)\, .
\end{equation}
This equation is one of the main results of our work, so we devote the next subsection to test it for different entangling regions, namely, the round ball, small deformations of it and thin strips.

\subsection{Explicit checks}
Let us perform some explicit verifications of our new formula for $F(A)$ in a few cases. On the one hand, we will explicitly evaluate the RT functional introducing a geometric regulator and extract $F(A)$ from the constant piece. On the other hand, we will directly evaluate  $\mathbf W_5\left(2\RT\right)$. Comparing both, we will find perfect agreement in all cases considered.

\subsubsection{Sphere}

As a first check, let us compute the renormalized area $\mathbf A_\Sigma^{\text{ren}}$ of the RT surface associated to a four-ball entangling region $\mathbb B^4$ and the generalized Willmore energy of its double-copied $2\RT$ surface 
 --- which turns out to be a round sphere $\mathbb S^4$ --- and see that they match. 

\begin{figure}
\centering
\parbox{7cm}{
\includegraphics[width=7cm]{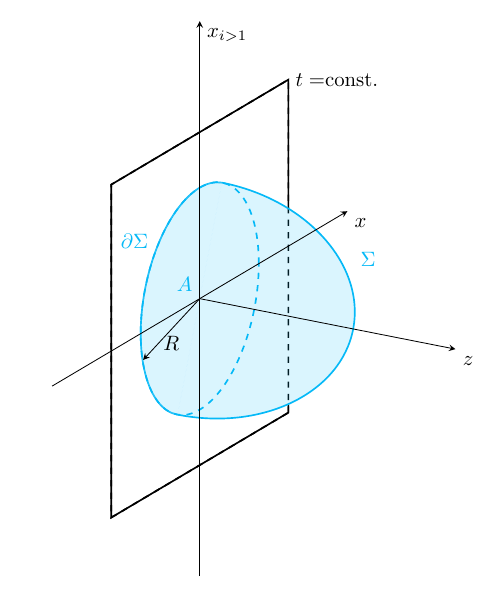}
\caption{Ball-shaped entangling region $A$ with radius $\rho=R$ and its cobordant ($\partial A=\partial \Sigma$) codimension-two surface $\Sigma$.}
\label{fig:1}}
\qquad
\begin{minipage}{7cm}
\includegraphics[width=7cm]{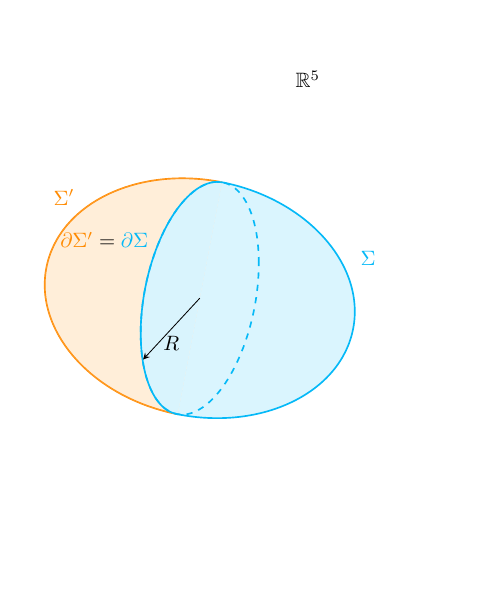}
\caption{Double-copied surfaces $\Sigma$ and $\Sigma'$ glued along the umbilical line defined by $\partial \Sigma=\partial \Sigma'$.}
\label{fig:2}
\end{minipage}
\end{figure}

The metric of the dual geometry is given by pure AdS$_6$ which, in Poincaré-AdS coordinates, reads
\begin{equation}\label{eq:PoincareAdS6}
    \diff s^2=g_{\alpha\beta}\diff x^\alpha\diff x^\beta=\frac{\Ls^2}{r^2\cos^2\theta}\left(\diff t^2+\diff r^2+r^2\diff \theta^2+r^2\sin^2\theta\diff\Omega^2_3\right)\, ,
\end{equation}
where $0\leq\theta\leq \frac{\pi}{2}$, with the conformal boundary located at $\theta=\frac{\pi}{2}$, and $\diff \Omega^2_3=\diff\theta_1^2+\sin^2\theta_1\left(\diff\theta_2^2+\sin^2\theta_2\diff\theta_3^2\right)$ represents the line element for the angular directions of the $\mathbb S^3$, with $0\leq\theta_1\leq\pi$, $0\leq\theta_2\leq\pi$ and $0\leq\theta_3\leq2\pi$. For such entangling region, the RT surface $\RT$ is given by the embedding 
\begin{equation}\label{eq:sphind}
    \RT:\left\{t=\text{const.}\, ,r=R\right\}\, ,\quad  \diff s^2_{\RT}=\gamma_{ab}^{\text{sp}}\diff y^a\diff y^b=\frac{\Ls^2}{\cos^2\theta}\left(\diff \theta^2+\sin^2\theta\diff\Omega^2_3\right)\, .
\end{equation}
The setup can be seen in Figure~\ref{fig:1}. Using this, it is a straightforward exercise to check that the bare area $\mathbf A\left(\RT\right)=\int_{\RT}\diff^4y\sqrt{\gamma}$ of the codimension-two RT surface, equipped with metric $\gamma_{ab}$, is given by \cite{Ryu:2006ef,Nishioka:2009un}
\begin{equation}\label{eq:renAsphbare}
    \mathbf A\left(\RT\right)=2\pi^2\Ls^4\int_0^{\frac{\pi}{2}-\frac{\delta}{R}}\sec\theta\tan^3\theta\diff\theta=\frac{2\pi^2\Ls^4}{3}\left(\frac{R}{\delta}\right)^3-\frac{5\pi^2\Ls^4}{3}\frac{R}{\delta}+\frac{4\pi^2\Ls^4}{3}+\mathcal{O}\left(\delta\right)\, ,
\end{equation}
where we introduced $\delta$ as a ultraviolet (UV) cutoff and $R$ is the ball radius at the conformal boundary, which is located at $\theta=\frac{\pi}{2}$. As anticipated in Eq.~\eqref{eq:Scases}, we obtain two divergent contributions: the so-called area-law term and a codimension-two divergent piece. Our expression for the renormalized area \eqref{eq:renA4} cancels these contributions, \emph{i.e.}, \cite{Anastasiou:2018rla}
\begin{equation}\label{eq:renAsph}
\mathbf A^{\text{ren}}\left(\RT\right)=\mathbf A\left(\RT\right)-\frac{\Ls^4}{24}\int_{\RT}\diff^4y\sqrt{\gamma}\,\mathcal X_4^{\RT}+\frac{4\pi^2\Ls^4}{3}\chi(\RT)=\frac{4\pi^2\Ls^4}{3}\,,
\end{equation}
where we used that the Euler characteristic of the RT surface is one, $\chi(\RT)=1$, as it is homeomorphic to a four-ball.\footnote{For an explicit cancellation of the divergences using the Chern form, see Appendix~\ref{app:sph}.} Notice that the boundary terms appearing in Eq.~\eqref{eq:renA4} vanish identically for this geometry, so we only have to consider the bulk quantities. Inspecting the formula, we observe that the first two terms in the second equality of Eq.~\eqref{eq:renAsph} cancel each other, because $\mathcal X_4^{\RT}=\frac{24}{\Ls^4}$ for the four-ball entangling region, leaving the topological piece as the only contribution to $\mathbf A^{\text{ren}}\left(\RT\right)$. In turn, this fact makes manifest the non-local nature of EE in odd-dimensional CFTs \cite{Anastasiou:2020smm}.

% \begin{figure}%
%     \centering
%     \subfloat[\centering Ball-shaped entangling region $A$ with radius $\rho=R$ and its cobordant $\partial A=\partial \Sigma$ codimension-two surface $\Sigma$.]{{\includegraphics[width=7cm]{figures/sphericalEE.pdf} }}%
%     \qquad
%    \subfloat[\centering Double-copied surfaces $\Sigma$ and $\Sigma'$ glued along the umbilical line defined by $\partial \Sigma=\partial \Sigma'$.]{{\includegraphics[width=7cm]{figures/sphericaldoubling.pdf} }}%
%    \caption{Spherical entangling surface.}%
%    \label{fig:sphere}%
% \end{figure}

Now, let us turn our attention to the expression of the generalized Willmore energy derived in Eq.~\eqref{eq:FG}. First, we need to rescale the six-dimensional background in Eq.~\eqref{eq:PoincareAdS6} with $\varphi=\log\frac{\Ls}{z}$ in Eq.~\eqref{eq:scaledg}. By doing so, the embedding of the rescaled RT surface reads
\begin{equation}\label{eq:resphere}
\diff \tilde s_{\RT}^2=\tilde\gamma_{ab}^{\text{sp}}\diff y^a\diff y^b=R^2\left(\diff \theta^2+\sin^2\theta\diff\Omega^2_3\right)\,,
\end{equation}
with the geometric quantities
\begin{equation}
    \tilde n^{(t)}_t=\tilde n^{(r)}_r=1\, ,\quad \tilde K^{(t)}{}_{ab}=0\, ,\quad \tilde K^{(r)}{}_{ab}=\frac{1}{R}\tilde\gamma_{ab}^{\text{sp}}\, ,
\end{equation}
which yields a particularly simple $\mathcal J_{\RT}=\frac{1}{R^4}$ in Eq.~\eqref{eq:J5}. Following the procedure discussed in Sec.~\ref{sec:GW4}, we have to consider a doubled-copied $2\RT=\RT\cup\RT'$ surface associated to our spherical entangling surface of radius $R$ and glue them along their boundaries $\partial \RT$ and $\partial \RT'$. The setup is described in Figure~\ref{fig:2}. By doing so, we find that the generalized Willmore energy for the doubled-copied RT surface yields
\begin{equation}
     \mathbf W_5\left(2\RT\right)=2\pi^2\int_0^{\frac{\pi}{2}}\diff\theta\, \sin^3\theta=\frac{8\pi^2}{3}\,,
\end{equation}
which, by means of relation \eqref{eq:FG}, reproduces the finite piece of the four-ball holographic EE 
\begin{equation}\label{eq:F0}
    F_0=\frac{\pi^2\Ls^4}{3\GN}\, ,
\end{equation}
where we denoted $F_0\equiv F\left(\mathbb B^4\right)$.

Now, to test further our results, let us turn our attention to an entangling surface with less symmetries.

\subsubsection{Small deformations of the sphere}\label{sec:defs}
Consider a four-ball entangling region like the one described in the previous section. However, now we are interested in studying small perturbations around such geometry --- we denote such entangling region as $\mathbb B^4_{\epsilon}$ --- in the angular direction $\theta_1$. In particular, let us consider infinitesimal deformations described by the polar equation 
\begin{equation}\label{eq:RTdef}    \rho\left(\theta_1\right)=R\left[1+\epsilon\sum_\ell a_\ell Y_\ell(\theta_1)+\mathcal{O}\left(\epsilon^2\right)\right]\,, \quad Y_\ell(\theta_1)=\frac{1}{2\pi^2\sqrt{\sin\theta_1}}Q_{\ell+\frac{1}{2}}^{\frac{1}{2}}\left(\cos\theta_1\right)\, ,
\end{equation}
where $a_\ell$ is a coefficient that controls the deformation and $Q_{n}^{m}(x)$ is an associated Legendre function of second kind. Such expression --- which can also be found in Ref.~\cite{Bueno:2015lza} --- corresponds to a subset of the deformations considered by Mezei in Ref.~\cite{Mezei:2014zla}. In the same reference, the embedding of the associated RT surface is provided, namely
\begin{align}
r(\theta,\theta_1)=&R\left[1+\epsilon \,\Theta(\theta,\theta_1)+\mathcal O\left(\epsilon^2\right)\right]\ , \\
\Theta(\theta,\theta_1)=&\sum_\ell a_\ell Y_\ell(\theta_1)\tan^\ell\frac{\theta}{2}\frac{1+(\ell+1)\cos\theta+\frac{\ell(\ell+2)}{3}\cos^2\theta}{1+\cos\theta}\, ,
\end{align}
where we use $r(\theta,\theta_1)$ to differentiate it from the coordinate describing the shape of the deformation of the entangling region, $\rho(\theta_1)$. In turn, the induced metric given by
\begin{equation}\label{eq:RTmetdef}
    \diff s^2_{\RT}=\gamma_{ab}^{\text{sp}}\diff y^a\diff y^b+\epsilon^2\frac{\Ls^2}{\cos^2\theta}\left({\Theta'}^2\diff \theta^2+{\dot \Theta}^2\diff \theta_1^2+2\Theta'\dot \Theta\diff\theta\diff\theta_1\right)+\mathcal{O}\left(\epsilon^3\right)\, ,
\end{equation}
where we introduced the shorthand notation $\Theta'=\partial_\theta \Theta$ and $\dot \Theta=\partial_{\theta_1} \Theta$. Using this expression, we can compute the bare area of the RT surface and obtain
\begin{align}\label{eq:eqnoninvder}
\mathbf A\left(\RT\right)&=\mathbf A^{\text{ren}}\left(\RT\right)+\frac{2\pi^2\Ls^4}{3}\left(\frac{R}{\delta}\right)^3-\frac{5\pi^2\Ls^4}{3}\frac{R}{\delta}\notag\\
&+\epsilon^2\frac{\ell(\ell+2)\Ls^4}{144\pi}\left[6\left(\frac{R}{\delta}\right)^3-\left(4\ell(\ell+2)-3\right)\frac{R}{\delta}\right]+\ldots,
\end{align}
where we included all the universal terms in $\mathbf A^{\text{ren}}\left(\RT\right)$. In Appendix~\ref{ap:defs}, we show that the nonuniversal terms in Eq.~\eqref{eq:eqnoninvder} are exactly cancelled by the boundary terms present in Eq.~\eqref{eq:renA4nc}. In turn, we isolate the universal terms as
\begin{equation}\label{eq:ArenDef}
\mathbf A^{\text{ren}}\left(\RT\right)=\frac{4\pi^2\Ls^4}{3}+\epsilon^2\frac{\Ls^4}{72\pi}\sum_\ell a_\ell^2\left(\ell-1\right)_{5}+\mathcal{O}\left(\epsilon^4\right)\, , \quad \text{where} \quad \left(x\right)_{n}\equiv \frac{\Gamma(x+n)}{\Gamma(x)}\, ,
\end{equation}
 is the Pochhammer symbol. %For the sake of clarity, we have omitted the nonuniversal in this expression. However, in appendix \ref{ap:defs}, we show that the these terms in the bare area, namely
%are exactly cancelled by the boundary terms present in \eqref{eq:renA4nc}

Proceeding similarly as before, we now compute the generalized Willmore energy and check if it coincides with the expression for the renormalized area. The rescaled metric reads
\begin{align}
    \diff\tilde s^2_{\RT}=&\tilde\gamma_{ab}^{\text{sp}}\diff y^a\diff y^b+2\epsilon \Theta\tilde\gamma_{ab}^{\text{sp}}\diff y^a\diff y^b\notag\\
    &+\epsilon^2\left[\Theta^2\tilde\gamma_{ab}^{\text{sp}}\diff y^a\diff y^b+R^2{\Theta'}^2\diff\theta^2+R^2{\dot \Theta}^2\diff\theta^2_1+2R^2\Theta'\dot \Theta\diff\theta\diff\theta_1\right]+\mathcal{O}\left(\epsilon^3\right)\, ,
\end{align}
where we employed the expression for $\tilde\gamma_{ab}^{\text{sp}}$ given in Eq.~\eqref{eq:resphere}. For this embedded surface, we can compute their normal vectors
\begin{align}
    n_r^{(r)}=&1-\frac{\epsilon^2}{2}\left({\Theta'}^2+\frac{\dot \Theta^2}{\sin^2\theta}\right)+\mathcal O\left(\epsilon^4\right)\,,\\
    n_{\theta}^{(r)}=-\epsilon R\Theta'&\sin\theta+\mathcal{O}\left(\epsilon^3\right)\, ,\quad n_{\theta_1}^{(r)}=- \epsilon R\dot \Theta\sin\theta+\mathcal{O}\left(\epsilon^3\right)\, ,
\end{align}
as well as the extrinsic curvatures that appear in the integrand, namely
\begin{align}
\tilde K^{(t)}{}_{ab}=&0\, ,\\
\tilde K^{(r)}{}_{ab}=&\frac{1}{R}\tilde\gamma_{ab}^{\text{sp}}+\epsilon R\left[\left(\Theta-\Theta''\right)\diff\theta^2+\left(\Theta\sin^2\theta-\ddot \Theta-\cos\theta\sin\theta\Theta'\right)\diff\theta_1^2\right.\nonumber\\
&+\left(\diff\theta_2^2+\sin^2\theta_2\diff\theta_3^2\right)\sin\theta_1\left(\sin\theta\sin\theta_1\left(\Theta\sin\theta-\Theta'\cos\theta\right)-\dot\Theta\cos\theta_1\right)\nonumber\\
&+2\left(\dot \Theta\cot\theta-\dot\Theta'\right)\diff\theta\diff\theta_1\big]-\frac{\epsilon^2R}{2}\left[\left(\dot\Theta^2\csc^2\theta-3{\Theta'}^2\right)\diff\theta^2+\left({\Theta'}^2\sin^2\theta-3{\dot\Theta}^2\right)\diff\theta_1^2\right.\notag\\
&\left.+\sin ^2\theta_1\left(\diff\theta_2^2+\sin^2\theta_2\diff\theta_3^2\right) \left(\dot\Theta^2+{\Theta'}^2\sin ^2\theta\right)-8\Theta'\dot\Theta\diff\theta\diff\theta_1\right]\, .
\end{align}
Using them for the three terms appearing in $\mathcal J_{\RT}$, we find that the generalized Willmore energy for the double-copied perturbed RT surface reads
\begin{equation}\label{eq:Gdeform}
\mathbf W_5\left(2\RT\right)=\frac{8\pi^2}{3}+\frac{\epsilon^2}{36\pi}\sum_\ell a_\ell^2\left(\ell-1\right)_{5}+\mathcal{O}\left(\epsilon^4\right)\, ,
\end{equation}
which is twice the renormalized area \eqref{eq:ArenDef} modulo the AdS scale (as expected) and
matches exactly the result of the holographic EE using the relation \eqref{eq:FG}, this is
\begin{equation}\label{eq:FGdeform}
F(\mathbb B^{d-1}_\epsilon)=\frac{\pi^2\Ls^4}{3\GN}+\epsilon^2\frac{\Ls^4}{288\pi\GN}\sum_\ell a_\ell^2\left(\ell-1\right)_{5}+\mathcal{O}\left(\epsilon^4\right)\, .
\end{equation}
This expression is also in agreement with Mezei's formula, as it should \cite{Mezei:2014zla} --- see also \cite{Allais:2014ata}. Indeed, in that paper it was pointed out that, when considering a slightly deformed spherical entangling region $\mathbb B^{d-1}_\epsilon$ in any dimension, the leading correction to the finite piece of holographic EE is controlled by the flat-space stress-tensor two-point function\footnote{The coefficient $C_{T}$ is a universal quantity, defined from  $\langle T_{\mu\nu}(x)T_{\rho\sigma}(0)\rangle_{\mathbb{R}^d}=\frac{C_{T}}{x^{2d}}\left[I_{\mu(\rho}I_{\sigma)\nu}-\frac{\delta_{\mu\nu}\delta_{\rho\sigma}}{d}\right]$, which holds for general CFTs, where $I_{\mu\nu} \equiv  \delta_{\mu\nu}-2\frac{x_\mu x_\nu}{x^2}$ is a theory-independent tensorial structure \cite{Osborn:1993cr}.} charge $C_{T}$ --- this holographic result was later shown to hold for arbitrary CFTs \cite{Faulkner:2015csl}. It is easy to check that Eq.~\eqref{eq:FGdeform}  can indeed be rewritten as
\begin{equation}\label{eq:FAdef}
   F(\mathbb B^{d-1}_\epsilon)=1+\epsilon^2\frac{\pi^3}{8640}C_T\sum_\ell a_\ell^2(\ell-1)_5+\mathcal O \left(\epsilon^4\right)\, ,
\end{equation}
where we used that for CFTs dual to Einstein gravity $C_T=\frac{30\Ls^4}{\pi^4\GN}$ \cite{Liu:1998bu}, and where both the functional dependence on the $a_\ell$, the $\ell$ and the overall coefficient match Mezei's general formula. %and we normalized by  the finite piece of holographic EE in the case of a spherical entangling region $F_0$, as given in \eqref{eq:F0}.

\subsubsection{Infinite strip}

\begin{figure}
\centering
\parbox{7cm}{
\includegraphics[width=7cm]{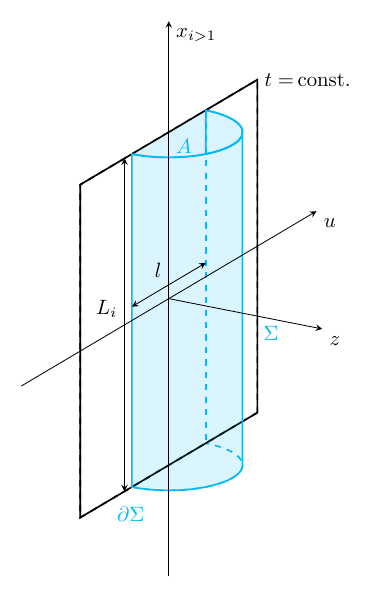}
\caption{Infinite strip entangling region $A$ of width $l$ in the $u$ direction. We introduced $L_i$ as IR regulators for the transversal directions $x_i$. The surface $\Sigma$ is cobordant ($\partial A=\partial \Sigma$), where $\partial \Sigma$ is composed by two parallel boundaries located at $u=\pm l/2$, i.e., $\partial \Sigma=\partial \Sigma^{l/2}\cup\partial \Sigma^{-l/2}$.
}
\label{fig:3}}
\qquad
\begin{minipage}{7cm}
\includegraphics[width=7cm]{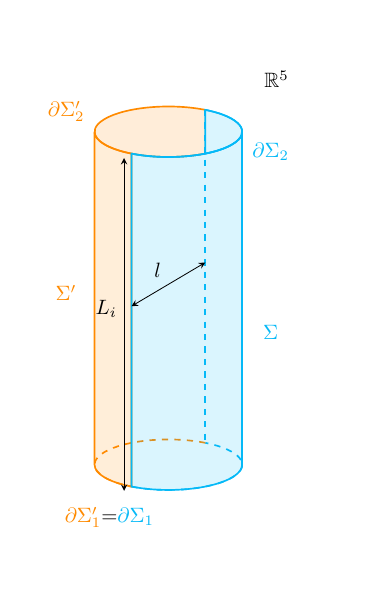}
\caption{Double-copied surfaces $\Sigma$ and $\Sigma'$ glued along the umbilical line defined by $\partial \Sigma_1=\partial \Sigma_1'$ located at $u=\pm l/2$. The introduction of the IR regulators $L_i$ defines a pair of symmetric boundaries $\partial \Sigma_2$ and $\partial \Sigma_2'$ pointing upwards and downwards along the transverse directions $x_i$.
}
\label{fig:4}
\end{minipage}
\end{figure}

As a final check, let us discuss the case of the infinite strip entangling region with width $l$ --- see Figure~\ref{fig:3}. In principle, this region --- as well as its associated RT surface --- is non-compact and the expression in terms of the generalized Willmore energy derived in Eq.~\eqref{eq:GW5} is not guaranteed to hold. However, we will see that it still captures the renormalized area of the RT surface and, as a consequence, the finite part of the holographic EE.

% \begin{figure}%
%     \centering
%     \subfloat[\centering Infinite strip entangling region $A$ of width $l$ in the $u$ direction. We introduced $L_i$ as IR regulators for the transversal directions. The dual geometry $\Sigma$ is cobordant $\partial A=\partial \Sigma$, where $\partial \Sigma$ composed by two parallel boundaries located at $u=\pm l/2$, this is $\partial \Sigma=\partial \Sigma^{l/2}\cup\partial \Sigma^{-l/2}$]{{\includegraphics[width=6cm]{figures/stripEE.pdf} }}%
%     \qquad
%     \subfloat[\centering Double-copied strip]{{\includegraphics[width=6cm]{figures/stripdoubling.pdf} }}%
%     \caption{2 Figures side by side}%
%     \label{fig:example}%
% \end{figure}

As usual, our starting point is Poincaré-AdS$_6$ spacetime \eqref{eq:PoincareAdS6} written, this time, in cylindrical coordinates --- which means performing the change of coordinates $z=r\cos\theta$, $u=r\sin\theta$, $x_1=r\sin\theta_1\cos\theta_2$, $x_2=r\sin\theta_1\sin\theta_2\cos\theta_3$ and $x_3=r\sin\theta_1\sin\theta_2\sin\theta_3$ ---, this is
\begin{equation}
\diff s^2=g_{\alpha\beta}\diff x^\alpha\diff x^\beta=\frac{\Ls^2}{z^2}\left(\diff u^2+\diff z^2+\diff\mathbf x^2\right)\, ,
\end{equation}
where $\diff\mathbf x^2=\diff x_1^2+\diff x_2^2+\diff x_3^2$. The codimension-two RT surface is defined through the embedding
\begin{equation}\label{eq:PAdS6cyl}
    \RT:\left\{t=\text{const.}\, , z=z(u)\right\}\, ,\quad  \diff s^2_{\RT}=\frac{\Ls^2}{z^2(u)}\left[\left(1+{z'}^2(u)\right)\diff u^2+\diff \mathbf x^2\right]\, ,
\end{equation}
with $z(u)$ such that the area is minimized and with the boundary condition $z(u=\pm l/2)=0$, which corresponds to the location of the conformal boundary. We can write the area functional and find a conserved quantity --- a sort of Hamiltonian --- associated to $u$ translations which then provides a first-order differential equation for $z(u)$, namely $z^8{z'}^2+z^8=z_\star^8$ --- here, $z_\star^8$ is related to the conserved quantity, and it represents the maximum value of $z$ that the surface reaches at $u = 0$. In turn, this allows us to change variables as
\begin{equation}\label{eq:firstI}
    \diff u=\frac{z^4}{\sqrt{z_\star^8-z^8}}\diff z\, .
\end{equation}
  Duplicating by symmetry the increasing branch $u\in\left[-l/2,0\right]$, the bare area can be computed --- after changing variables using Eq.~\eqref{eq:firstI} --- as \cite{Ryu:2006ef}
\begin{equation}\label{eq:Abarestrip}
    \mathbf A\left(\RT\right)=2L_i^3\Ls^4z_\star^4\int_\delta^{z_\star}\frac{\diff z}{z^4\sqrt{z_\star^8-z^8}}=\frac{2\Ls^4}{3}\left(\frac{L_i}{\delta}\right)^3-\frac{2\sqrt{\pi}\Gamma(5/8)\Ls^4}{3\Gamma(1/8)}\left(\frac{L_i}{z_\star}\right)^3\, ,
\end{equation}
where, again, we introduced $\delta$ as an UV cut-off for this entangling region. From this, we can easily extract the universal coefficient as the second term in the RHS. Note that the maximum depth $z_\star$ can be written in terms $l$ using the relation \cite{Ryu:2006ef}
\begin{equation}\label{eq:zl}
    \frac{l}{2}=\int_0^{z_\star}\frac{z^4\diff z}{\sqrt{z_\star^8-z^8}}=\frac{\sqrt{\pi}\Gamma(5/8)}{\Gamma(1/8)}z_\star\, ,
\end{equation}
which we will use afterwards.

Unlike in the case of the sphere, the expression for the renormalized area \eqref{eq:renA4} is not applicable to the infinite strip as it is. The reason is that in our argument we assumed that the entangling region is compact and, as a consequence, we used the Gauss-Bonnet theorem to exchange the Chern form $\mathcal B_3$ with the four-dimensional Euler characteristic and density, $\chi(\RT)$ and $\mathcal X_4^{\RT}$ respectively. This is not the case for the infinite strip, for which we introduced IR regulators $L_i$'s to characterize the region. As such, we cannot expect the Euler characteristic, which is a topological quantity, to carry information regarding IR regulators in our setup. Because of this, in order to find the renormalized area of the RT surface, we undo the Gauss-Bonnet theorem to return to a formula for the renormalized area which does not require compact codimension-two manifolds. In Appendix \ref{ap:strip} we show the explicit cancellation of divergences, so that we obtain
\begin{equation}\label{eq:Arenstrip}
 \mathbf A^{\text{ren}}\left(\RT\right)=-\frac{16\pi^2\Ls^4}{3}\left(\frac{\Gamma(5/8)}{\Gamma(1/8)}\right)^4\left(\frac{L_i}{l}\right)^3\, ,
\end{equation}
where we have used Eq.~\eqref{eq:zl} to express the result in terms of the width $l$ of the strip. The result agrees with the one obtained from direct subtraction of the area-law divergence in Eq.~\eqref{eq:Abarestrip}, as it should.

Now, let us turn our attention to generalized Willmore energy \eqref{eq:GW5}. First, we need the geometric quantities associated to the flat background --- which, as usual, amounts to rescaling the metric \eqref{eq:PAdS6cyl} as \eqref{eq:scaledg} with $\varphi=\log (\Ls/z)$ being the conformal factor ---, this is
\begin{equation}
\diff \tilde s^2_{\RT}=\tilde g_{ab}\diff x^a\diff x^b= \frac{z_\star^8}{z^8} \diff u^2+\diff \mathbf x^2\, ,
\end{equation}
where we have already used the relation \eqref{eq:firstI} to express the induced metric of $\RT$ in terms of $z_\star$. From here, we obtain
\begin{equation}\label{eq:rsstrip}
    \tilde n_t^{(t)}=1\,, \quad \tilde n_u^{(z)}=-\frac{\sqrt{z_\star^8-z^8}}{z_\star^4}\, ,\quad \tilde n_z^{(z)}=\frac{z^4}{z_\star^4}\, ,\quad \tilde K^{(z)}{}_{uu}=\frac{4z_\star^4}{z^5}\, ,\quad \tilde K^{(z)}=\frac{4z^3}{z_\star^4}\, ,
\end{equation}
with every other component of $\tilde K_{ab}$ identically vanishing. Taking into account these considerations and using the quantities in Eq.~\eqref{eq:rsstrip}, we obtain the generalized Willmore energy of the RT surface as
\begin{equation}
\mathbf W_5\left(2\RT\right)=4L_i^3\int_0^{z_\star}\frac{3 z^4 \left(z_\star^8-2 z^8\right)\diff z}{z_\star^{12} \sqrt{z_\star^8-z^8}}=-\frac{32\pi^2}{3}\left(\frac{\Gamma(5/8)}{\Gamma(1/8)}\right)^4\left(\frac{L_i}{l}\right)^3\, ,
\end{equation}
where we have again introduced $L_i$ as IR regulators and expressed the result in terms of the width $l$ using Eq.~\eqref{eq:zl}. Once again we see that following the relation between the generalized Willmore energy and the finite part of holographic EE \eqref{eq:FG} we obtain the expected result \cite{Ryu:2006ef}
\begin{equation}\label{stripi}
F(A)=-\frac{4\pi^2\Ls^4}{3\GN}\left(\frac{\Gamma(5/8)}{\Gamma(1/8)}\right)^4\left(\frac{L_i}{l}\right)^3\, .
\end{equation}

\section{$F(A)$ has no global bounds for $d=5$ CFTs}\label{sec4}

As discussed in Sec.~\ref{sec:2}, the disk, $\mathbb B^2$, minimizes the finite part of holographic EE in the vacuum state among all possible entangling regions for three-dimensional holographic CFTs. This can be seen from the saturation of the lower bound of the Willmore energy $\mathbf W_3$, which occurs when the double-copied submanifold $2\RT$ is a sphere. This result extends to arbitrary CFTs, as shown in Ref.~\cite{Bueno:2021fxb} and summarized in Eq.~\eqref{fa0}.

Let us now exploit our new formula in terms of the generalized Willmore energy $\mathbf W_5$ to explore the shape dependence of $F(A)$ for five-dimensional holographic theories. 
%Let us now address the situation for EE for holographic CFT$_5$ and its relation to generalized Willmore energy. 
Of course, Mezei's formula --- of which Eq.~\eqref{eq:FAdef} is a particular case --- implies that the higher-dimensional version of the disk-like entangling region, $\mathbb B^4$, is a local minimum of $F(A)$ for small derformations of the ball not just for holographic CFTs, but for completely general CFTs \cite{Mezei:2014zla}. Hence, any small deformation away from the round ball will produce an increase of $F(A)$. 

An obvious question is then whether $A=\mathbb B^4$ is a global minimum. It is immediate to see that this is not the case. This follows from the result for the strip region \eqref{stripi}, which implies that the holographic EE is not bounded \emph{from below} in five dimensions \cite{Ryu:2006ef}. Namely, $F(A)$ takes arbitrarily negative values as the IR regulators are made arbitrarily large. This is somewhat suprising. Indeed, as shown in Eq.~\eqref{eq:Scases}, the EE of a general odd-dimensional CFT contains a universal term of the form 
\begin{equation}
    S_{\rm EE}(A) \supset (-1)^{\frac{d-1}{2}}F(A)\, .
\end{equation}
With this normalization, $F(A)$ is such that it takes a positive value for a round ball region $\mathbb{B}^{d-1}$ for general $d$-dimensional CFTs --- see \emph{e.g.}, \cite{Casini:2011kv,Klebanov:2011gs,Nishioka:2009un}. On the other hand, in the case of an infinite strip, $F(A)$ is positive for $d=3,7,11,\dots$ but negative for $d=5,9,13,\dots$ This is the case not only for holographic theories \cite{Nishioka:2009un}, but also for free scalars and fermions \cite{Casini:2009sr} and, presumably, for general CFTs. Hence, we immediately learn that $F(A)$ is unbounded from above for $d=3,7,11,\dots$ and from below for $d=5,9,13,\dots$. Hence, while for $d=7,11,\dots$ it is still plausible that ---just like for $d=3$ --- $\mathbb{B}^{d-1}$ is a global minimum, for  $d=5,9,\dots$ this is not the case. Indeed, in those cases there must exist families of entangling regions which interpolate between round balls and very thin ``strip-like'' regions such that $F(A)$ starts growing as we depart from the ball, it reaches a maximum for certain region, it takes a value coincident with the ball one for some other region, it vanishes for some other as we keep deforming, and then it takes increasingly negative values as the strip shape is approached. On the other hand, it is still possible that $F(A)$ is bounded above in those cases, although this seems unlikely.

In order to gain further insight on this matter, let us now consider smooth non-perturbative deformations of $\mathbb B^4$ and feed them to our newly constructed functional $\mathbf W_5$.
%The idea is to consider a $\mathbb B^4$ entangling region and smooth non-perturbative deformations thereof to check if we can recover the result for the strip and whether there is an upper bound on $\mathbf G_\Sigma$. 
 For concreteness, let us consider a four-dimensional ellipsoidal surface embedded in $\mathbb R^5$, described by the equation $R^2=\sum_{i=1}^5\frac{x_i^2}{b_i^2}$ in Cartesian coordinates and where $b_i$ represent the length of each of the semiaxes. In spherical coordinates, this reads\footnote{For clarity, we choose different angular variables with respect to Sec.~\ref{sec:defs}. In this case their range is $0\leq\theta_i\leq\pi$ for $i=1,2,3$ and $0\leq\phi\leq2\pi$.}
\begin{align}
R^2&=\frac{r^2\cos^2 \theta_1}{b_1^2}+\frac{r^2\sin^2\theta_1\cos^2 \theta_2}{b_2^2}+\frac{r^2\sin^2\theta_1\sin^2 \theta_2\cos^2 \theta_3}{b_3^2}\notag\\
&+\frac{r^2\sin^2\theta_1\sin^2 \theta_2\sin^2 \theta_3\cos^2\phi}{b_4^2}+\frac{r^2\sin^2\theta_1\sin^2 \theta_2\sin^2 \theta_3\sin^2\phi}{b_5^2}.
\end{align}
We consider two families of ellipsoids: a first one $\Sigma_{a}^{(1)}$ with axes of length $\left(b_1,b_2,b_3,b_4,b_5\right)=\left(1,1,1,1,a\right)$ and a second one $\Sigma_{a}^{(2)}$ with $\left(1,1,1,a,a\right)$. For them, we evaluate the generalized Willmore energy $\mathbf W_5\left(\Sigma_{a}\right)$ and normalize it by the value of the four-sphere $\mathbf W_5\left(\mathbb S^4\right)=\frac{8\pi^2}{3}$, with the latter retrieved from the former when we take $a=1$ in $\Sigma_{a}^{(1)}$ and $\Sigma_{a}^{(2)}$ respectively. Notice that the ellipsoids spanned by the embeddings $\Sigma_{a}^{(1)}$ and $\Sigma_{a}^{(2)}$ do not necessarily correspond to double-copied RT surfaces as we run the parameter $a$. However, from this exercise we can get an intuition about what to expect for RT surfaces and, hence, for holographic EE.

We observe that the ratio $\mathbf W_5\left(\Sigma_{a}\right)/\mathbf W_5\left(\mathbb S^4\right)$ for the first ellipsoid $\left(1,1,1,1,a\right)$ can be computed analytically, obtaining
{\small\begin{equation}
    \frac{\mathbf W_5\left(\Sigma_{a}^{(1)}\right)}{\mathbf W_5\left(\mathbb S^4\right)}=\frac{315 \left(15 a^2-16\right) a^8 \text{arcsec}\, a+\sqrt{a^2-1} \left(10613 a^8-7778 a^6-2376 a^4-16 a^2-128\right)}{17920 a^6 \left(a^2-1\right)^{3/2}}\, .
\end{equation}}
There are two important limits that can be derived from this expression, namely when $a$ is small and large. They correspond to geometries tending to $\mathbb B^4$ and $\mathbb S^3\times \mathbb R$, respectively. In both regimes we see that generalized Willmore energy grows indefinitely
\begin{equation}\label{eq:GWalim}
\mathbf W_5\left(\Sigma_{a\ll1}^{(1)}\right)=\frac{1}{140a^6}+\mathcal{O}\left(a^{-4}\right)\, ,\quad \mathbf W_5\left(\Sigma_{a\gg1}^{(1)}\right)=\frac{135\pi}{1024}a+\mathcal{O}\left(a^{0}\right)\, .
\end{equation}
This behavior was previously reported numerically in Ref.~\cite{Graham:2017bew}, where the authors suggested that $\mathbf W_5\left(\Sigma_{a}^{(1)}\right)$ is a convex function. Here, based on the analytical expression \eqref{eq:GWalim} we can unequivocally check that this is the case. Incidentally, this means that $F(A)$ should not be expected to possess an upper bound either for general $d=5$ CFTs.

\begin{figure}
\centering

\includegraphics[width=11cm]{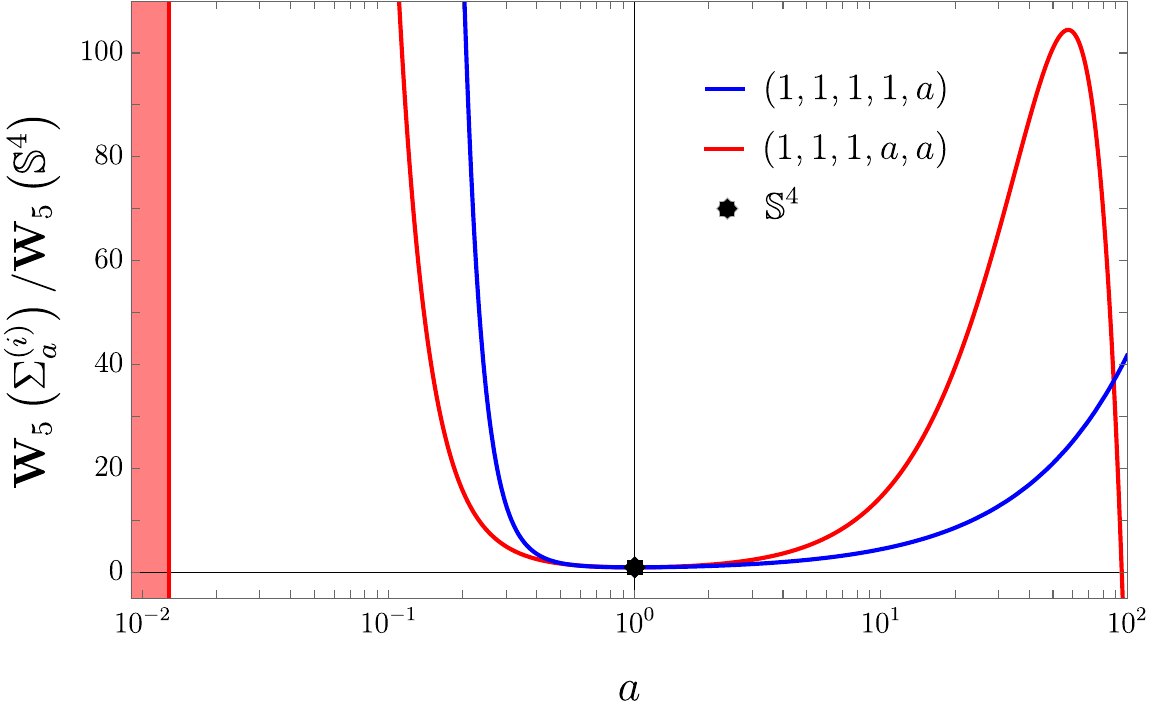}
\caption{Linear-logarithmic plot of $\mathbf W_5\left(\Sigma_{a}^{(i)}\right)/ \mathbf W_5\left(\mathbb S^4\right)$ with $i=1,2$, representing the $(1,1,1,1,a)$ (blue line) and $(1,1,1,a,a)$ (red line) ellipsoids respectively for different values of $a$. The values represented for the former are analytical while the latter are numerical. The red region for small values of $a$ indicates that the $(1,1,1,a,a)$ ellipsoid oscillates wildly between $-\infty$ and $+\infty$ in this regime. We also include the point $a=1$ corresponding to a round sphere $\mathbb S^4$ and represented by an eight-pointed star. As we can see, while $\mathbb S^4$ --- corresponding to an $A=\mathbb B^4$ holographic entangling region --- is a local minimum of the generalized Willmore functional, this is neither bounded from below nor from above for general regions.} 
\label{fig:5}
\end{figure}

Regarding the second ellipsoid $\left(1,1,1,a,a\right)$, we are not able to find an analytical expression for $\mathbf W_5\left(\Sigma_{a}^{(2)}\right)/\mathbf W_5\left(\mathbb S^4\right)$. However, we can still find numerical results running for different values of $a$. In Figure~\ref{fig:5} we plot these results as well as including the analytical expression for the $\left(1,1,1,1,a\right)$ ellipsoid, \eqref{eq:GWalim}. As expected, for both ellipsoids, $\mathbf W_5\left(\Sigma_{a}\right)$ in the regime $a\rightarrow1$ tends to the value of the sphere $\mathbf W_5\left(\mathbb S^4\right)$ and corresponds to a local minimum. In the regime $a\ll1$,\footnote{This limit is the closest one to the thin strip case considered throughout the paper.} we observe that the ratio $\mathbf W_5\left(\Sigma_{a}^{(2)}\right)/\mathbf W_5\left(\mathbb S^4\right)$ oscillates wildly between $-\infty$ and $+\infty$. We represented this oscillation with a red region in the figure. On the other hand, the $a\gg1$ regime, which can be associated to $\mathbb R^2\times \mathbb S^2$ geometry is unbounded from below. An analogous behavior was previously reported in Ref.~\cite{martino2023duality}.\footnote{One could ask what is the situation with other ellipsoids, such as $\left(1,1,a,a,a\right)$, $\left(1,a,a,a,a\right)$. They are connected to the two cases studied so far by means of the duality $a\leftrightarrow1/a$ and, hence, their large and small $a$ regimes are respectively exchanged.}

In summary, in this simple setup we observe that the generalized Willmore energy \eqref{eq:GW5}: i) has a local minimum for $\mathbb S^4$, corresponding to an entangling region $\mathbb B^4$; ii) it is neither bounded from below nor from above. From the holographic EE point of view, i) was previously known as it follows from Mezei's formula. On the other hand, ii) reveals that $\mathbf W_5$ and, consequently, $F(A)$ for five-dimensional CFTs can take arbitrarily negative and positive values for certain entangling regions. 

\section{Conclusions}\label{sec:4}
In this paper we have presented a new formula for the vacuum EE universal term $F(A)$ for holographic theories dual to Einstein gravity in five (boundary) dimensions. The resulting expression generalizes the Willmore energy functional which captures the corresponding result in the three-dimensional case. This generalized Willmore energy, $\mathbf W_5$, is given --- in agreement with previous results in the mathematical literature \cite{Graham:2017bew,zhang2017grahamwittens} ---  by an integral over the doubled RT surface embedded in $\mathbb{R}^5$ of a linear combination of terms of order-$4$ in extrinsic curvatures --- see Eq.~\eqref{ww5}. As we have seen, in contradistinction to the three-dimensional case, $\mathbf W_5$ is both unbounded from above and from below, which implies the same conclusion for $F(A)$ at least in the holographic case. A more detailed scrutiny of the free-field results available in the literature strongly suggests that this is a general feature of five-dimensional CFTs.

$\mathbf W_5$ was obtained here from the evaluation of six-dimensional LLP CG \cite{Maldacena:2011mk} in the conically singular orbifold defined through the LM procedure  \cite{Lewkowycz:2013nqa} following the prescription by Miao given in Ref.~\cite{Miao:2015iba}. By requiring the resulting functional to be a conformal invariant we were able to derive $\mathbf W_5$ as well as the so-called reduced Hawking mass. The latter followed from imposing the bulk manifold to be an Einstein space, whereas $\mathbf W_5$ was obtained from further imposing the surface to be extremal and the bulk to be pure AdS.

There are some future directions which would be worth exploring. On the one hand, it is natural to wonder about the universal term in the holographic EE for Einstein gravity in seven (boundary) dimensions, which would yield yet another generalized Willmore energy functional, $\mathbf{W}_7$. This should involve some linear combination of terms of order 6 in extrinsic curvatures of the doubled RT surface embedded in $\mathbb{R}^7$. Presumably, this should follow from a procedure analogous to the one exploited here, involving this time certain eight-dimensional CG. 

On a different front, it is also natural to explore generalizations of the original Willmore energy $\mathbf{W}_3$ motivated by holographic EE. Indeed, considering higher-curvature terms in the gravitational action modifies the RT formula introducing corrections to the area functional. Consequently, the corresponding universal term which in the case of Einstein gravity is given by $\mathbf{W}_3$ will be modified by terms of higher order in extrinsic curvatures of the doubled RT functional. The obvious first case to consider is the one of quadratic gravities \cite{Fursaev:2013fta}, from which one would expect corrections to $\mathbf{W}_3$ involving terms of order $4$ in extrinsic curvatures. Aside from the interpretation of the resulting functionals in the context of holographic EE, this procedure could be used to obtain somewhat canonical higher-curvature generalizations of $\mathbf{W}_3$ which may be of interest from a mathematical perspective. 

Furthermore, the proposed generalization of the reduced Hawking mass in four dimensions \eqref{eq:RHM5} opens the possibility to derive new holographic EE bounds for states different from the CFT vacuum as well. The two-dimensional reduced Hawking mass demonstrates a monotonous behavior under inverse mean-curvature flows that gives rise to a generic bound for holographic EE in three-dimensional CFTs~\cite{Fischetti:2016fbh}. Our new four-dimensional reduced Hawking mass involves desirable terms such as the Gauss-Bonnet density, which is expected to follow a monotonous behavior under a flow which is not necessarily the inverse mean-curvature one. It is worth noting that, as the reduced Hawking mass renormalizes the area of arbitrary (\emph{i.e.}, not necessarily minimal) surfaces, it would yield the finite part of holographic EE even for quantum extremal surfaces, which take into account quantum corrections in the entropy due to the bulk degrees of freedom \cite{Faulkner:2013ana,Engelhardt:2014gca} and $\mathcal{O}\left(1/N\right)$ corrections as well.

Finally, the results presented here show that the $\mathbb B^4$ EE in the case of holographic  five-dimensional CFTs is a somewhat less significant quantity than the $\mathbb B^2$ one in the three-dimensional counterpart. Indeed, while in the latter case it provides a universal lower bound for $F(A)$ for general CFTs, in the former it only does so for small deformations around the ball region. It would be interesting to explore the consequences of this fact in light of putative generalizations to five dimensions of the three-dimensional conformal bounds presented in Ref.~\cite{Bueno:2023gey}.

\section*{Acknowledgements}
We thank Dorian Martino, Erik Tonni, Andrew Waldron and Matteo Zatti for useful discussions. The work of GA is funded by ANID FONDECYT grants No. 11240059 and 1240043. The work of IJA is funded by ANID FONDECYT grants No.~11230419 and~1231133. IJA is grateful to PB and the ICC at the Universitat de Barcelona for their hospitality. PB was supported by a Ramón y Cajal fellowship (RYC2020-028756-I), by a Proyecto de Consolidaci\'on Investigadora (CNS 2023-143822) from Spain's Ministry of Science, Innovation and Universities, and by the grant PID2022-136224NB-C22, funded by MCIN/AEI/ 10.13039/501100011033/FEDER, UE . The work of JM is supported by ANID FONDECYT Postdoctorado Grant No. 3230626. RO was supported by Anillo Grant ACT210100 \emph{Holography and its applications to High Energy Physics, Quantum Gravity and Condensed Matter Systems} and ANID FONDECYT Regular grants No. 1230492, 1231779, 1240043 and 1240955. AVL is supported by the F.R.S.-FNRS Belgium through conventions FRFC PDRT.1025.14 and IISN 4.4503.15, as well as by funds from the Solvay Family. AVL thanks PB and ICCUB for their hospitality during the early stages of this project.

\appendix
\section{Notation and conventions}\label{conventions}

In this appendix we present the conventions used throughout the paper. In the first column of Table \ref{tab:notation}, we provide a list of objects defined on the different manifolds presented in the first line. The gravity theory is defined on the $(d+1)$-dimensional bulk manifold $\mathcal M$ and its dual CFT lives on its boundary $\partial \mathcal M$. We denote $\Sigma$ as the codimension-two manifold in which the RT surface associated to the entangling region $A$ is defined and $\partial \Sigma=\partial A$ as its boundary. In Table \ref{tab:notation}, we also differentiate between various embeddings that can be defined for submanifolds, such as $\partial \Sigma$, which can be embedded in either $\Sigma$ or $\partial \mathcal M$.

\begin{table}[h!]
    \centering
    \begin{tabular}{|l|c|c|c|c|c|}
     \cline{2-6}
    \multicolumn{1}{c|}{}
 & $\mathcal{M}$ &  $\partial\mathcal{M} \subset \mathcal{M}$ & $\Sigma \subset \mathcal{M}$ & $\partial \Sigma \subset \Sigma$ & $\partial A = \partial \Sigma \subset \partial \mathcal{M}$ \\ \hline\hline
    Indices & $\alpha,\ldots,\lambda$ &$\mu,\ldots,\omega$& $a,\ldots,h$ & $i,\ldots,q$ & $i,\ldots,q$ \\
    Coordinates & $x^\alpha$ &$X^\mu$& $y^a$ & $Y^i$ & $Y^i$ \\
    Metric & $g_{\alpha\beta}$ & $h_{\mu\nu}$ & $\gamma_{ab}$ & $\sigma_{ij}$ & $\sigma_{ij}$ \\
    Covariant derivative & $\nabla_\alpha$ & $\nabla_\mu^{\partial \mathcal M}$ & $\nabla_a^{\Sigma}$ & $\nabla_i^{\partial \Sigma}$ & $\nabla_i^{\partial \Sigma}$ \\
    Riemann tensor & $R_{\alpha\beta\gamma\delta}$ & $r_{\mu\nu\rho\sigma}$ & $\mathcal R_{abcd}$ & $\mathfrak R_{ijkl}$ & $\mathfrak R_{ijkl}$ \\
    Unit normal(s) &  & $n_{\alpha}$ & $N^{A}{}_{\alpha}$ & $\mathfrak{n}_{a}$ & $\mathfrak l^{I}{}_{\mu}$ \\
    Extrinsic curvature & & $k_{\mu\nu}$ &$K^{A}{}_{ab}$& $\mathfrak K_{ij}$ & $\kappa^{I}{}_{ij}$ \\\hline
    \end{tabular}
    \caption{Notation and conventions}
    \label{tab:notation}
\end{table}

Assuming an embedding $x^{\alpha}=x^{\alpha} \left(y^a\right)$, we define the projection vielbein to $\Sigma$, $\gamma^\alpha_a =\frac{\partial x^{\alpha}}{\partial y^{a}}$. Similar constructions are valid for the other submanifolds in the table above. Unit normals are taken to be outward-pointing (for the codimension-two cases, $\Sigma \subset \mathcal{M}$ and $\partial \Sigma \subset \partial \mathcal{M}$, we introduce indices $A, B, \dots$ and $I, J, \dots$, respectively, labelling the two normals). From this definition, extrinsic curvatures are obtained as
\begin{equation}
K^{A}{}_{ab} = \gamma^{\alpha}_a \gamma^{\beta}_b \nabla_\beta N^{A}{}_{\alpha} \, .
\end{equation}
In addition, we can write the contraction of $K^{A}{}_{ab}$ with bulk indices as
\begin{equation}\label{eq:covK}
    K^{\gamma}{}_{\alpha\beta}=\gamma_{\alpha}^a\gamma_{\beta}^bN_A{}^\gamma K^{A}{}_{ab}\, .
\end{equation}
It is important to note that in the codimension-two decomposition that we perform here, the normal bundle indices play the role of labels for objects residing in the hypersurface. This becomes evident when the Gauss-Codazzi relations for the Christoffel symbol are considered. In particular, these can be cast as
\begin{equation}
\Gamma _{aA}^{B} =0 \,,\quad \Gamma _{a A}^{b} = -K_{Aa}{}^{b} \,, \quad \Gamma _{ab}^{A} =K^{A}{}_{ab} \,.
\label{GaussCodChrist}
\end{equation}
Notice that the normal bundle index $A$ of the Christoffel is interpreted as the label indicating the direction along which the extrinsic curvature is computed. On the other hand, since the $\partial \Sigma \subset \Sigma$ embedding is performed only along the radial direction, the label can be omitted and the corresponding extrinsic curvature can be written as
\begin{equation}
\mathfrak{K}_{ij} = \sigma^{a}_i \sigma^{b}_j \nabla_{b}^{\Sigma} \mathfrak{n}_{a} \, .
\end{equation}
We will occasionally make use of partial contractions of indices along normal or tangent directions to a given submanifold. Let us exemplify this with $\Sigma \subset \mathcal{M}$. Given that we take the normals to satisfy $g^{\alpha \beta} N^A{}_{\alpha} N^B{}_{\beta} = \delta^{AB}$ (in Euclidean signature), we can decompose the metric into tangent and normal components as
\begin{equation}
g_{\alpha \beta} = \gamma_{\alpha \beta} + \delta_{A B} N^A{}_{\alpha} N^B{}_{\beta} \, .
\end{equation}
We then abbreviate normal contractions using indices $A, B, \dots$, e.g.,
\begin{equation}
R^{A}_A = R^{\alpha \beta} n^A{}_{\alpha} n^B{}_{\beta} \delta_{AB} \, ,
\end{equation}
and similarly for tangent ones,
\begin{equation}
R^{a}_a = R^{\alpha \beta} \gamma_{\alpha \beta} \, .
\end{equation}
In the case in which we choose coordinates adapted to the surface, so that $x^{\alpha} = (x^A, x^a = y^a)$ and $N^A{}_{\alpha} = \delta^A_{\alpha}$, the previous expressions reduce to the contraction of normal / tangent indices.

% Thus, the extrinsic curvature of the codimension-two submanifold $\Sigma$ along the direction $\alpha$ of the bulk manifold $\mathcal M$, given by the normal vector $n^\alpha_a$, can be written as
% \begin{equation}
% K^\alpha{}_{ab} = \gamma^\beta_a \gamma^\delta_b \nabla_\beta n^\alpha_\delta \,.
% \end{equation}
% Similarly, for the embedding $X^{\mu}=X^{\mu} \left(Y^i\right)$, the projection vielbein to $\partial \Sigma$ is defined as 
% $\sigma^\mu_i = \frac{\partial X^\mu}{\partial Y^i}$. In this case, we define the extrinsic curvature of $\partial \Sigma$ along the direction $a$ of $\partial \mathcal M$, given by the normal vector $n^\mu_i$, as
% \begin{equation}
% \mathfrak K^\mu{}_{ij}=\sigma^\nu_i\sigma^\rho_j \nabla_\nu^{\partial \mathcal M}n^\mu_\rho \,,
% \end{equation}
% where $\nabla^{\partial \mathcal M}$ is the covariant derivative of the boundary manifold $\partial \mathcal M$.

Throughout the text we make use of the so-called Schouten tensor, which is defined in general dimension $D$ as
\begin{equation}
    S_\alpha^\beta\equiv \frac{1}{(D-1)}\left(R_\alpha^\beta-\frac{1}{2D}R\delta_\alpha^\beta\right)\, .
\end{equation}
The Cotton tensor is in turn defined from this as
\begin{equation}
    C_{\alpha\beta\gamma}\equiv \nabla_\gamma S_{\alpha\beta}-\nabla_\beta S_{\alpha\gamma}\, .
\end{equation}
Also, the Weyl tensor can be defined using the Schouten tensor as
\begin{equation}\label{eq:WSdef}
W_{\alpha\beta}^{\gamma\delta}\equiv  R_{\alpha\beta}^{\gamma\delta}-4S_{[\alpha}^{[\gamma}\delta_{\beta]}^{\delta]}\, .
\end{equation}
%which can be taken as the definition of the Weyl.
Finally, the Bach tensor is defined as
\begin{equation}
B_{\alpha\beta}\equiv S_{\gamma\delta}W_\alpha{}^\gamma{}_\beta{}^\delta+2\nabla^\delta\nabla_{[\delta} S_{\alpha]\beta} \, .
\end{equation}

\section{Conformal covariantization of Einstein-AdS gravity}\label{confcomple}

In order to perform the conformal covariantization procedure, we are seeking to construct Weyl-invariant scalar densities $\mathcal{I}$, namely
\begin{equation}
\updelta_{\varphi}\mathcal{I} =0 \,,
\label{scalardensity}
\end{equation}
where $\varphi$ is the local scaling parameter of the metric as in Eq.~\eqref{eq:scaledg}.
%, \emph{i.e.},
%\begin{equation}
%g_{\mu \nu}\rightarrow %\tilde{g}_{\mu \nu} %=e^{2\sigma} g_{\mu \nu} \,.
%\end{equation}
For an infinitesimal Weyl transformation, the metric behaves as
\begin{equation}
\updelta_{\varphi}g_{\alpha\beta} =2\varphi g_{\alpha\beta} \,. \label{metric}
\end{equation}
Based on this relation, we determine the behavior of the Ricci scalar and Schouten tensor, as
\begin{align}\updelta_{\varphi}R &= -2\varphi R -2\left (D -1\right ) \dal \varphi \, , \label{Ricci} \\
\updelta_{\varphi}S_{\alpha\beta} &= - \nabla _\alpha \nabla _\beta\varphi  \,. \label{Schouten}
\end{align}
From this, it follows the Weyl tensor invariance,
\begin{equation}
\updelta_{\varphi}W^{\delta}{}_{\gamma\alpha\beta}=0\,.
\label{Weyl}
\end{equation}
Additional expressions that will be useful for our computations are
\begin{align}
\updelta_{\varphi}C_{\alpha \beta \gamma } &= -W_{\delta \alpha \beta \gamma } \nabla ^{\delta } \varphi  \,, \label{Cotton} \\
\updelta_{\varphi}B_{\alpha \beta } &= -2\sigma B_{\alpha \beta } +\left (D -4\right )\left (C_{\alpha \beta \gamma } +C_{\beta \alpha \gamma }\right ) \nabla ^{\gamma } \varphi  \,, \label{Bach}
\end{align}
where $C_{\alpha \beta \gamma}$ and $B_{\alpha \beta }$ are the Cotton and Bach tensors, respectively. Then, it is straightforward to show the Weyl invariance of the Pfaffian of the Weyl~\eqref{WE3}.\footnote{The Pfaffian of a certain tensor $X_{\alpha\beta}^{\gamma\delta}$ in even $D$ dimensions is given by $\text{pf}(X)\equiv\delta ^{\gamma_{1}\delta_{1}\ldots\gamma_{D/2}\delta_{D/2}}_{\alpha_{1}\beta_{1}\ldots\alpha_{D/2}\beta_{D/2}}X^{\alpha_{1}\beta_{1}}_{\gamma_{1}\delta_{1}}\cdots X^{\alpha_{D/2}\beta_{D/2}}_{\gamma_{D/2}\delta_{D/2}}$.} Indeed, we get that
\begin{align}
\updelta_{\varphi}\left (\sqrt{|g|}\mathcal Y_6\right )&=\sqrt{|g|} \delta _{\beta _{1}\ldots \beta _{6}}^{\alpha _{1}\ldots \alpha _{6}} \left[\frac{1}{2}W_{\alpha _{1}\alpha _{2}}^{\beta _{1}\beta _{2}}W_{\alpha _{3}\alpha _{4}}^{\beta _{3}\beta _{4}}W_{\alpha _{5}\alpha _{6}}^{\beta _{5}\beta _{6}}\left (g^{ -1}\updelta_{\varphi}g\right )
+3W_{\alpha _{1}\alpha _{2}}^{\beta _{1}\beta _{2}}W_{\alpha _{3}\alpha _{4}}^{\beta _{3}\beta _{4}}\updelta_{\varphi}W_{\alpha _{5}\alpha _{6}}^{\beta _{5}\beta _{6}}\right] \nonumber  \\
&=\sqrt{|g|}\delta _{\beta _{1}\ldots \beta _{6}}^{\alpha _{1}\ldots \alpha _{6}} 6 \varphi \left (W_{\alpha _{1}\alpha _{2}}^{\beta _{1}\beta _{2}}W_{\alpha _{3}\alpha _{4}}^{\beta _{3}\beta _{4}}W_{\alpha _{5}\alpha _{6}}^{\beta _{5}\beta _{6}} - W_{\alpha _{1}\alpha _{2}}^{\beta _{1}\beta _{2}}W_{\alpha _{3}\alpha _{4}}^{\beta _{3}\beta _{4}}W_{\alpha _{5}\alpha _{6}}^{\beta _{5}\beta _{6}}\right ) =0 \,,
\end{align}
due to the fact that $\updelta_{\varphi}W_{\gamma \delta }^{\alpha \beta } = -2\varphi W_{\gamma \delta }^{\alpha \beta }$. On the other hand, the term $\mathcal Y_4$ breaks Weyl invariance explicitly. In particular, for the corresponding scalar density, we obtain
\begin{align}
&\updelta_{\varphi}\left (\sqrt{|g|}\delta _{\beta _{1}\ldots \beta _{5}}^{\alpha _{1}\ldots \alpha _{5}}W_{\alpha _{1}\alpha _{2}}^{\beta _{1}\beta _{2}}W_{\alpha _{3}\alpha _{4}}^{\beta _{3}\beta _{4}}S_{\alpha _{5}}^{\beta _{5}}\right ) \nonumber \\
&=\sqrt{|g|}\delta _{\beta _{1}\ldots \beta _{5}}^{\alpha _{1}\ldots \alpha _{5}} \left[\frac{1}{2}W_{\alpha _{1}\alpha _{2}}^{\beta _{1}\beta _{2}}W_{\alpha _{3}\alpha _{4}}^{\beta _{3}\beta _{4}}S_{\alpha _{5}}^{\beta _{5}}\left (g^{ -1}\updelta_{\varphi}g\right )
+2W_{\alpha _{1}\alpha _{2}}^{\beta _{1}\beta _{2}}\left (\updelta_{\varphi}W_{\alpha _{3}\alpha _{4}}^{\beta _{3}\beta _{4}}\right )S_{\alpha _{5}}^{\beta _{5}} +W_{\alpha _{1}\alpha _{2}}^{\beta _{1}\beta _{2}}W_{\alpha _{3}\alpha _{4}}^{\beta _{3}\beta _{4}}\updelta_{\varphi}S_{\alpha _{5}}^{\beta _{5}}\right] \nonumber \\
&= -\sqrt{|g|}\delta _{\beta _{1}\ldots \beta _{5}}^{\alpha _{1}\ldots \alpha _{5}}W_{\alpha _{1}\alpha _{2}}^{\beta _{1}\beta _{2}}W_{\alpha _{3}\alpha _{4}}^{\beta _{3}\beta _{4}} \nabla ^{\beta _{5}} \nabla _{\alpha _{5}} \varphi  \,.
\end{align}
Thus, we are seeking compensating terms that will render this expression invariant under infinitesimal Weyl transformations. This is  achieved by rewriting the last term as the Weyl variation of a scalar density. To do so, after integrating by parts, the latter can be cast in the form
\begin{align}
\updelta_{\varphi}&\left (\sqrt{|g|}\delta _{\beta _{1}\ldots \beta _{5}}^{\alpha _{1}\ldots \alpha _{5}}W_{\alpha _{1}\alpha _{2}}^{\beta _{1}\beta _{2}}W_{\alpha _{3}\alpha _{4}}^{\beta _{3}\beta _{4}}S_{\alpha _{5}}^{\beta _{5}}\right )
= -32\sqrt{|g|}W_{\gamma \delta }^{\alpha \beta }C_{\alpha }^{\gamma \delta } \nabla _{\beta } \varphi  -4 \nabla ^{\gamma }\left (\sqrt{|g|}W^2 \nabla _{\gamma } \varphi \right ) \nonumber  \\
&+16 \nabla ^{\gamma }\left (\sqrt{|g|}W^2\updelta_{\varphi}C_{\alpha \beta }^{\delta }\right ) +32 \nabla ^{\gamma }\left (\sqrt{|g|} \varphi W^2C_{\alpha \beta }^{\delta }\right ) \,,
\label{deltaL61}
\end{align}
where the relation
\begin{equation}
W_{\alpha \beta }^{\gamma \delta } \nabla _{\delta }\varphi  =\updelta_{\varphi}C_{\alpha \beta }^{\gamma } +2 \varphi C_{\alpha \beta }^{\gamma } \,, \label{cotton2}
\end{equation}
was used.
%Similarly, one can show that\begin{equation}W_{\nu \lambda }^{\mu \rho } \nabla ^{\lambda }\sigma  =\updelta_{\varphi}C_{\nu }^{\mu \rho } +4\sigma C_{\nu }^{\mu \rho } . \label{cotton3}
%\end{equation}
Furthermore, the Weyl variation of the divergence of a generic vector field $V_\alpha$, reads
\begin{align}
\updelta_{\varphi}\left (\sqrt{|g|} \nabla ^{\alpha}V_{\alpha}\right )
=\sqrt{|g|} \nabla ^{\alpha }\left (4 \varphi V_{\alpha } +\updelta_{\varphi}V_{\alpha }\right ) \,.
\end{align}
This property allows us to simplify the expression of Eq.~\eqref{deltaL61}. Indeed, we write
\begin{equation}
\updelta_{\varphi}\left (\sqrt{|g|} \nabla ^2W^2\right ) =\sqrt{|g|} \nabla ^{\alpha }\left [4 \varphi  \nabla _{\alpha }W^2 +\updelta_{\varphi}\left(\nabla _{\alpha }W^2\right )\right ]= -4\sqrt{|g|} \nabla ^{\alpha }\left (W^2 \nabla _{\alpha } \varphi \right ) \,,
\end{equation}
where we denoted $\nabla^2=\nabla_\alpha\nabla^\alpha$.  Equivalently, we can write
\begin{equation}
\nabla ^{\alpha}\left (\sqrt{|g|}W^2 \nabla _{\alpha} \varphi \right ) = -\frac{1}{4}\updelta_{\varphi}\left (\sqrt{|g|} \nabla^2W^2\right )\, .
\label{weylvar1}
\end{equation}
On top of that, the following relation is valid
\begin{align}
\updelta_{\varphi}\left [\sqrt{|g|} \nabla ^{\gamma }\left (W_{\delta \gamma }^{\alpha \beta }C_{\alpha \beta }^{\delta }\right )\right ] &=\sqrt{|g|} \nabla ^{\gamma }\left [4 \varphi W_{\delta \gamma }^{\alpha \beta }C_{\alpha \beta }^{\delta } +\updelta_{\varphi}\left (W_{\delta \gamma }^{\alpha \beta }C_{\alpha \beta }^{\delta }\right )\right ] \nonumber  \\
&=\sqrt{|g|} \nabla ^{\gamma }\left (2 \varphi W_{\delta \gamma }^{\alpha \beta }C_{\alpha \beta }^{\delta } +W_{\delta \gamma }^{\alpha \beta }\updelta_{\varphi}C_{\alpha \beta }^{\delta }\right ) \,.
\end{align}
After some algebraic manipulation the latter can be cast in the form
\begin{equation}
\nabla ^{\mu }\left (\sqrt{|g|}W_{\nu \mu }^{\alpha \beta }\updelta_{\varphi}C_{\alpha \beta }^{\nu }\right ) =\updelta_{\varphi}\left [\sqrt{|g|} \nabla ^{\mu }\left (W_{\nu \mu }^{\alpha \beta }C_{\alpha \beta }^{\nu }\right )\right ] -2 \nabla ^{\mu }\left (\sqrt{|g|} \varphi W_{\nu \mu }^{\alpha \beta }C_{\alpha \beta }^{\nu }\right )\,.
\label{weylvar2}
\end{equation}
Replacing Eqs.~(\ref{weylvar1},\ref{weylvar2}) into Eq.~\eqref{deltaL61}, we get
\begin{align}
\updelta_{\varphi}\left (\sqrt{|g|}\delta _{\beta _{1}\ldots \beta _{5}}^{\alpha _{1}\ldots \alpha _{5}}W_{\alpha _{1}\alpha _{2}}^{\beta _{1}\beta _{2}}W_{\alpha _{3}\alpha _{4}}^{\beta _{3}\beta _{4}}S_{\alpha _{5}}^{\beta _{5}}\right ) =& -32\sqrt{|g|}C_{\gamma }^{\alpha \beta }(\updelta_{\varphi}C_{\alpha \beta }^{\gamma } +2 \varphi C_{\alpha \beta }^{\gamma }) \notag\\
&-2\updelta_{\varphi} \nabla ^{\alpha }\left [\sqrt{|g|}(8W_{\alpha \beta }^{\gamma \delta }C_{\gamma \delta }^{\beta } -W_{\kappa \beta }^{\gamma \delta } \nabla _{\alpha }W_{\gamma \delta }^{\kappa \beta })\right ] \,. \label{deltaL62}
\end{align}
As a final step, we have to Weyl-covariantize the remaining terms involving the Cotton squared contribution. For this term, we consider that
\begin{align}
\updelta_{\varphi}\left (\sqrt{|g|}C^2\right ) &=\frac{\sqrt{|g|}}{2}C^2\left (g^{ -1}\updelta_{\varphi}g\right ) +\sqrt{|g|}\left (C_{\gamma }^{\alpha \beta }\updelta_{\varphi}C_{\alpha \beta }^{\gamma } +C_{\alpha \beta }^{\gamma }\updelta_{\varphi}C_{\gamma }^{\alpha \beta }\right ) \\
&=2\sqrt{|g|}C_{\gamma }^{\alpha \beta }\left (\updelta_{\varphi}C_{\alpha \beta }^{\gamma } +2 \varphi C_{\alpha \beta }^{\gamma }\right ) \,.
\end{align}
Substituting this expression in Eq.~\eqref{deltaL62}, we obtain
\begin{equation}
\updelta_{\varphi}\left (\sqrt{|g|}\delta _{\beta _{1}\ldots \beta _{5}}^{\alpha _{1}\ldots \alpha _{5}}W_{\alpha _{1}\alpha _{2}}^{\beta _{1}\beta _{2}}W_{\alpha _{3}\alpha _{4}}^{\beta _{3}\beta _{4}}S_{\alpha _{5}}^{\beta _{5}}\right )
= -2\updelta_{\varphi}\left[\sqrt{|g|}\left(8C^2 + \nabla ^{\alpha }\hat J\right)\right] \,,
\end{equation}
where we denoted $\hat{J}^\alpha\equiv8 W^{\alpha\gamma\delta\beta}C_{\gamma\lambda\beta} - W^{\gamma\delta}_{\beta\varepsilon}\nabla^\alpha W^{\beta\varepsilon}_{\gamma\delta}$. As a consequence, the scalar density $\mathcal I_{4}$ is Weyl invariant, \emph{i.e.}, $\updelta_{\varphi}\mathcal I_{4} =0$, where
\begin{equation}
\mathcal I_{4} =\sqrt{|g|}\left (\frac{1}{2}\delta _{\beta _{1}\ldots \beta _{5}}^{\alpha _{1}\ldots \alpha _{5}}W_{\alpha _{1}\alpha _{2}}^{\beta _{1}\beta _{2}}W_{\alpha _{3}\alpha _{4}}^{\beta _{3}\beta _{4}}S_{\alpha _{5}}^{\beta _{5}}+8C^2 + \nabla ^{\alpha }\hat J_\alpha\right ) \,,
\end{equation}
which corresponds to the conformal covariantization, or Weyl completion, of the $-\frac{1}{2\Ls^2} \mathcal Y_4$ combination of Eq.~\eqref{eq:Y4cc}.

\section{Computation of $\Upsilon_{a b}$}\label{Appendixcomputation}
Our starting point on the derivation of the term $\Upsilon_{ab} $ in Eq.~\eqref{upsilonfin}, will be its covariant form
{\small\begin{equation} 
 \Upsilon_{ab} =\frac{1}{4}\gamma_a^{\alpha}\gamma_b^\beta\left[\frac{1}{16}\left (\nabla_{\alpha }K^{\gamma } \nabla_{\beta }K_{\gamma} - K^{\gamma }K^{\delta} R_{\alpha \gamma \beta \delta }\right ) +S_{\alpha \gamma }S_{\beta }^{\gamma} -B_{\alpha \beta }-S_{\gamma(\alpha} \nabla_{\beta)}K^{\gamma}-\frac{1}{2}K^{\gamma} \nabla_{\gamma} S_{\alpha \beta }\right]\,,
\end{equation}}
given in Ref.~\cite{Miao:2015iba}. After performing the integration by parts of the last term and, on parallel, expressing the Riemann tensor in terms of the Weyl and the Schouten tensors using expression \eqref{eq:WSdef}, the last expression yields the form
\begin{align}
\Upsilon_{ab}&=\frac{1}{4}\gamma_a^{\alpha}\gamma_b^{\beta}\bigg[\frac{1}{16}\big( \nabla_{\alpha }K^{\gamma } \nabla_{\beta }K_{\gamma } -K^{\gamma }K^{\delta }W_{\alpha \gamma \beta \delta } -K^{\gamma }K_{\gamma}S_{\alpha \beta } -K^{\gamma }K^{\delta }S_{\gamma \delta }g_{\alpha \beta } +2K^{\gamma }K_{(\alpha}S_{\beta) \gamma }\big) \nonumber  \\
&+S_{\alpha \gamma }S_{\beta }^{\gamma } -B_{\alpha \beta }-\frac{1}{2}\left (S_{\alpha \gamma } \nabla _{\beta }K^{\gamma } +K^{\gamma }C_{\beta \alpha \gamma } + \nabla _{\alpha }\left (K^{\gamma }S_{\beta \gamma }\right )\right )\bigg]\, ,
\end{align}
where $\gamma_{ab}$ is the intrinsic metric of $\Sigma$ and $\gamma_a^\alpha$ is the projector. At this point we can drop the covariant notation adopting the normal decomposition. In this case, only the extrinsic curvatures along the normal bundle directions survive, since the vector $n_{a}$ is tangent to $\Sigma$, leading to $K_{a} =0$. As a consequence, the last formula can be cast in the following form
\begin{gather}
\Upsilon_{ab} = \frac{1}{4}\left[\frac{1}{16}\big( \nabla _{a}K^{A} \nabla _{b}K_{A} -K^{A}K^{B}W_{a A b B} -K^{A}K_{A} S_{ab}-K^{A}K^{B}S_{AB}\gamma_{ab}\big)\right.\nonumber  \\
+\left.S_{a\alpha}S_{b}^{\alpha} -B_{ab} -\frac{1}{2}\left(S_{aA} \nabla_{b}K^{A} +K^{A}C_{abA} +\gamma_{b}^{\beta} \nabla_{a}\left(K^{A} S_{\beta A}\right )\right)\right] \,.
\end{gather}
The Gauss-Codazzi relations~\eqref{GaussCodChrist} allow us to express bulk covariant derivatives in terms of the covariant derivative $\nabla_a^{\Sigma}$, that is compatible with the induced metric $\gamma_{ab}$. As a consequence, the following term reads
\begin{equation}
 \nabla_{a}K^{A} \nabla _{b}K_{A} =\nabla_{a}^\Sigma K^{A}\nabla_{b}^\Sigma K_{A} +K^{A}K_{A}{}_a^{c}K_{B}K^{B}{}_{bc} \,.  
\end{equation}
On top of that, the next term where an explicit derivative appears, can be rewritten as 
% \begin{gather}
% S_{\bar{\kappa} \nu } \nabla _{\bar{\lambda}}\mathcal{K}^{\nu } =S_{\bar{\kappa} \nu }\left ( \partial _{\bar{\lambda}}\mathcal{K}^{\nu } +\Gamma _{\bar{\lambda}\mu }^{\nu }\mathcal{K}^{\mu }\right ) \nonumber  \\
% =S_{\bar{\kappa} \nu } \partial _{\bar{\lambda}}\mathcal{K}^{\nu } +S_{\bar{\kappa}\nu }\Gamma _{\bar{\lambda}\mu }^{\nu}\mathcal{K}^{\mu } \nonumber  \\
% =S_{\bar{\kappa}\nu } \partial _{\bar{\lambda}}\mathcal{K}^{\nu } +S_{\bar{\kappa}\nu }\Gamma _{\bar{\lambda}\left (i\right )}^{\nu }\mathcal{K}^{\left (i\right )} \nonumber  \\
% =S_{\bar{\kappa}\left (i\right )} \partial _{\bar{\lambda}}\mathcal{K}^{\left (i\right )} +S_{\bar{\kappa} \bar{\sigma}}\Gamma _{\bar{\lambda}\left (i\right )}^{\bar{\sigma}}\mathcal{K}^{\left (i\right )} \nonumber  \\
% =S_{\bar{\kappa} \left (i\right )}D_{\bar{\lambda}}\mathcal{K}^{\left (i\right )} -S_{\bar{\kappa}\bar{\sigma}}\mathcal{K}_{\left (i\right)\bar{\lambda}}^{\bar{\sigma}}\mathcal{K}^{\left (i\right )} \,.
% \end{gather}
\begin{equation}
S_{aA} \nabla _{b}K^{A}=S_{aA}\nabla_{b}^\Sigma K^{A} -S_a^c K^{A}{}_{bc}K_{A} \,.
\end{equation}
Finally, the last derivative contribution  can be analyzed as follows
% \begin{gather}
% \sigma _{\bar{\lambda}}^{\nu } \nabla _{\bar{\kappa}}\left (\mathcal{K}^{\mu }S_{\nu \mu }\right ) =\sigma _{\bar{\lambda}}^{\nu }\left [ \partial _{\bar{\kappa}}\left (\mathcal{K}^{\mu }S_{\nu \mu }\right ) -\Gamma _{\nu \bar{\kappa}}^{\mu }\mathcal{K}^{\sigma }S_{\sigma \mu }\right ] \nonumber  \\
% =\sigma _{\bar{\lambda}}^{\nu } \partial _{\bar{\kappa}}\left (\mathcal{K}^{\left (i\right )}S_{\nu \left (i\right )}\right ) -\Gamma_{\bar{\kappa}\bar{\lambda}}^{\mu }\mathcal{K}^{\left (i\right )}S_{\left (i\right )\mu } \nonumber  \\
% =\sigma _{\bar{\lambda}}^{\nu } \partial _{\bar{\kappa}}\left (\mathcal{K}^{\left (i\right )}S_{\nu \left (i\right )}\right ) -\Gamma _{\bar{\kappa}\bar{\lambda}}^{\bar{\sigma}}\mathcal{K}^{\left (i\right )}S_{\left (i\right )\bar{\sigma}} -\Gamma _{\bar{\kappa}\bar{\lambda}}^{\left (j\right )}\mathcal{K}^{\left (i\right )}S_{\left (i\right )\left (j\right )} \nonumber  \\
% =D_{\bar{\kappa}}\left (\mathcal{K}^{\left (i\right )}S_{\left (i\right )\bar{\lambda}}\right ) -\mathcal{K}_{\bar{\kappa}\bar{\lambda}}^{\left (j\right )}\mathcal{K}^{\left (i\right )}S_{\left (i\right )\left (j\right )} \,.
% \end{gather}
\begin{equation}
\gamma_{b}^{\beta} \nabla_{a}\left(K^{A} S_{\beta A}\right )=\nabla_{a}^\Sigma\left(K^{A}S_{Ab}\right) -K^{A}K^{B}{}_{ab}S_{AB} \,.
\end{equation}
Summing up all the previous contributions, the quantity $\Upsilon_{ab}$ now reads
\begin{align}
\Upsilon_{a b} &=\frac{1}{4}\left[\frac{1}{16}\left(\partial _{a}K^{A} \partial _{b}K_{A} +K_{A}K^{A}{}_{ac}K_{B}K^{B}{}_b^c -K^{A}K^{B} W_{aAbB}-K^{A}K_{A}S_{a b} \right.\right.\nonumber  \\
&\left.-K^{A}K^{B}S_{AB}\gamma_{a b}\right) +S_{a\alpha}S_{b}^{\alpha} -B_{a b}-\frac{1}{2}\left(S_{aA} \partial _{b}K^{A} -S_a^cK^{A}{}_{bc}K_{A} +K^{A}C_{a bA}\right. \nonumber  \\
&\left.\left.+\nabla_{a}^{\Sigma}\left (K^{A}S_{Ab}\right ) -K^{A}K^{B}{}_{ab}S_{AB} \right)\right]\,.
\end{align}

\section{Cancellation of divergences}\label{appdiv}

In this appendix we show explicitly the cancellation of divergences for the renormalized area in the case of different entangling regions. 

\subsection{Sphere}\label{app:sph}

Although it is easy to check that for the spherical entangling region \eqref{eq:renAsph} yields the renormalized area, for the sake of completeness here we show that the expression with the Chern form \eqref{eq:renA4nc} also does the job. Starting with the embedded metric of the RT surface associated to the spherical entangling region \eqref{eq:sphind}, it is easy to check that the boundary of this metric at $\theta=\pi/2-\delta/R$ is given by
\begin{equation}
    \diff s^2_{\partial \RT}=\sigma_{ij}^{\text{sph}}\diff Y^i\diff Y^j=\Ls^2\left[-\frac{2}{3}+\left(\frac{R}{\delta}\right)^2+\frac{1}{15}\left(\frac{\delta}{R}\right)^2+\mathcal{O}\left(\delta^4\right)\right]\diff\Omega_3^2\, .
\end{equation}
For this geometry we obtain
\begin{align}
\mathcal B_3^{\RT}&=-2\sqrt{\sigma}\left[2(\mathfrak{K}\mathfrak{R}-2\mathfrak{K}_i^j\mathfrak{R}_j^i)-\frac{2}{3}\left(\mathfrak{K}^3-3\mathfrak{K}\mathfrak{K}_i^j\mathfrak{K}_j^i-2\mathfrak{K}_i^j\mathfrak{K}_j^k\mathfrak{K}_k^i\right)\right]\\
&=8\sin^2\theta_1\sin\theta_2\left(\frac{R}{\delta}\right)^3-20\sin^2\theta_1\sin\theta_2\frac{R}{\delta}+\mathcal{O}\left(\delta^1\right)\, 
\end{align}
where we used that
\begin{equation}
\mathfrak{R}_i^j=\frac{2}{\Ls^2}\left(\frac{\delta}{R}\right)^2\delta_i^j+\mathcal{O}\left(\delta^4\right)\, ,\quad \mathfrak{K}_i^j=\frac{1}{\Ls}\left[1+\frac{1}{2}\left(\frac{\delta}{R}\right)^2\right]\delta_i^j+\mathcal{O}\left(\delta^4\right)\, .
\end{equation}
The last part of the boundary term in Eq.~\eqref{eq:renA4nc} vanishes identically $\mathcal K_{\partial \RT}=0$, because
\begin{equation}
    w_{ij}^{ij}=0\, , \quad \kappa^{I}{}_{\langle ij\rangle}=0\, .
\end{equation}
As a consequence, we see that
\begin{align}
    \mathbf A\left(\RT\right)^{\text{ren}}&=\mathbf A\left(\RT\right)-\frac{\Ls^4}{24}\int_{\partial\RT}\diff^3Y\mathcal B_3^{\RT}\\
    &=\mathbf A\left(\RT\right)-\frac{2\pi^2\Ls^4}{3}\left(\frac{R}{\delta}\right)^3+\frac{5\pi^2\Ls^4}{3}\frac{R}{\delta}+\mathcal{O}\left(\delta\right)\, ,\label{eq:Arensphere}
\end{align}
where the two terms carrying the UV regulator in this expression precisely cancel those appearing in the bare area \eqref{eq:sphind}.

\subsection{Small deformation of the sphere}\label{ap:defs}

Now, let us show explicitly that expression \eqref{eq:renA4nc} also achieves cancellation of divergences appearing in the area of the RT surface associated to the slightly deformed entangling region \eqref{eq:RTdef}. Starting with the induced metric of the RT surface \eqref{eq:RTmetdef}, we find the induced metric at the conformal boundary $\theta\rightarrow\pi/2$, finding
\begin{align}
\diff s^2_{\partial \RT}&=\sigma_{ij}^{\text{sph}}\diff Y^i\diff Y^j+\Ls^2{Y_\ell'}^2 \epsilon ^2 \left[\left(\frac{R}{\delta }\right)^2+\frac{1}{3} (1-\ell (\ell+2))\right.\notag\\
&\left.+\frac{1}{45}(5 (\ell-1) \ell (\ell+2) (\ell+3)+3)\left(\frac{\delta}{R}\right)^2\right]\diff\theta_1^2+\ldots\,,
\end{align}
where again, we have introduced an UV regulator $\delta$. The Chern form $\mathcal B_3^{\partial \RT}$ for this metric reads
% \begin{align}
% \mathcal B_3^{\partial \RT}&=-\frac{8}{\Ls^3}+\frac{4}{3\Ls^3}\left(\frac{\delta}{R}\right)^2\left[9+\epsilon ^2 \left(\ell^2 (\ell+2)^2 Y_\ell^2+2 \ell (\ell+2) Y_\ell \left(2Y_\ell' \cot \theta_1  +Y_\ell''\right)\right.\right.\notag\\
% &\left.\left.+6 Y_\ell'\csc ^2\theta_1 \left(Y_\ell'\cos 2\theta_1+Y_\ell''\sin2 \theta_1 \right)\right)\right]+\ldots
% \end{align}
\begin{align}
\mathcal B_3^{\partial \RT}=&8\sin^2\theta_1\sin\theta_2\left(\frac{R}{\delta}\right)^3-20 \sin^2 \theta_1 \sin\theta_2\frac{R}{\delta}+\epsilon^2\left[4\sin ^2\theta_1 \sin\theta_2 {Y_\ell'}^2\left(\frac{R}{\delta}\right)^3\right.\notag\\
+&\left(2\sin \theta_2 \left(2 l^2 (l+2)^2 \sin ^2\theta_1 Y_\ell^2+Y_\ell'
   \left(\left((2 l (l+2)-15) \sin ^2\theta_1+12\right) Y_\ell'\right.\right.\right.\notag\\
   +&\left.\left.\left.24 \sin \theta_1 \cos \theta_1
   Y_\ell''\right)+4 l (l+2) \sin \theta_1 Y_\ell \left(\sin \theta_1
   Y_\ell''+2 \cos \theta_1 Y_\ell'\right)\right)\right)\frac{R}{3 \delta }\Bigg]+\ldots
\end{align}
Finally, we need the partial trace of the Weyl tensor at the conformal boundary and the quadratic contraction of traceless extrinsic curvature of $\partial \Sigma$ embedded in $\partial \mathcal M$. The first one vanishes identically for our metric under consideration ($w_{ij}^{ij}=0$) whereas the second reads
\begin{equation}
\mathfrak{\kappa}^{I}{}_{\langle ij\rangle}\mathfrak{\kappa}_{I}{}^{\langle ij\rangle}=\frac{2}{3\Ls^2}\left(\frac{\delta}{R}\right)^2\left(Y_\ell''-Y_l'\cot\theta_1\right)^2\, ,    
\end{equation}
Putting all terms together in Eq.~\eqref{eq:renA4nc}, we find 
\begin{align}
\mathbf A^\text{ren}\left(\RT\right)&=\mathbf A\left(\RT\right)-\frac{2 \pi ^2 \Ls^4}{3} \left(\frac{R}{\delta}\right)^3+\frac{5 \pi ^2 \Ls^4}{3}\left(\frac{R}{\delta}\right)\notag\\
&-\epsilon^2\frac{\ell(\ell+2)\Ls^4}{144\pi}\left[6\left(\frac{R}{\delta}\right)^3-\left(4\ell(\ell+2)-3\right)\frac{R}{\delta}\right]+\ldots\, ,
\end{align}
which precisely cancels the divergences appearing in Eq.~\eqref{eq:eqnoninvder}, at quadratic order in $\epsilon$.

\subsection{Infinite strip}\label{ap:strip}

Let us show that the divergent piece in Eq.~\eqref{eq:Abarestrip} is cancelled when using expression \eqref{eq:renA4nc}. The starting point is the metric of the RT surface associated to the infinite strip entangling region \eqref{eq:PAdS6cyl}, which, after changing variables using Eq.~\eqref{eq:firstI} reads
\begin{equation}
    \diff s^2_{\RT}=\frac{L^2}{z^2}\left(\frac{z_\star^8\diff z}{z_\star^8-z^8}+\diff \mathbf{x}^2\right)\, .
\end{equation}
The conformal boundary of this geometry is located at $z=\delta$, and the induced metric at this locus is just flat space with a conformal factor, \emph{i.e.},
\begin{equation}\label{eq:BRTstrip}
    \diff s^2_{\partial\RT}=\frac{\Ls^2}{\delta^2}\diff \mathbf{x}^2\, .
\end{equation}
Using this, we can compute the quantities appearing in Eq.~\eqref{eq:renA4nc}, namely
\begin{equation}
    \mathcal B_3^{\RT}=\frac{8}{\delta^3}\, ,\quad  w_{ij}^{ij}=0\, ,\quad \kappa^I{}_{ij}=0,,
\end{equation}
where, for the second Chern form of the RT surface $\mathcal B_3^{\RT}$, we have used that for the conformally flat induced metric \eqref{eq:BRTstrip}, we have 
\begin{equation}
    \mathfrak R_{ij}^{kl}=0\, ,\quad \mathfrak{K}{}_{\langle ij\rangle}=\frac{1}{\Ls}\sigma_{ij}\, .
\end{equation}
Taking this into account, we obtain the renormalized area of the infinite strip following equation \eqref{eq:renA4nc}, this is
\begin{equation}\label{eq:Arenstrip}
    \mathbf A^{\text{ren}}\left(\RT\right)=\mathbf A\left(\RT\right)-\frac{\Ls^4}{24}\int_{\partial\RT}\diff^3Y\mathcal B_3^{\RT}=\mathbf A\left(\RT\right)-\frac{2\Ls^4}{3}\left(\frac{L_i}{\delta}\right)^3\, ,
\end{equation}
where we used that for the infinite strip we have two parallel boundaries, one located at $u=l/2$, say $\RT^{l/2}$, and another symmetric one located at $u=-l/2$, \emph{i.e.}, $\RT^{-l/2}$ ---see Figure~\ref{fig:3}. Thus, in the boundary term, there is a factor of two as they contribute symmetrically, i.e.,  $\partial\RT=\partial\RT^{l/2}\cup\partial\RT^{-l/2}$. With these considerations, we observe that the additional term appearing in Eq.~\eqref{eq:Arenstrip} cancels the divergence coming from the bare area of the RT surface \eqref{eq:Abarestrip}.

\bibliographystyle{JHEP}
\bibliography{CGWillmore.bib}

\end{document}